\documentclass[aps,prd,twocolumn,showpacs,eqsecnum,amsmath,amssymb,nofootinbib,superscriptaddress]{revtex4-1}

\usepackage{dcolumn}
\usepackage{bm}
\usepackage{graphics}
\usepackage{graphicx}
\usepackage{epsfig}
\usepackage{booktabs,multirow}
\usepackage{hhline}
\usepackage{slashed}
\usepackage{rccol}
\usepackage{mathrsfs}
\usepackage{enumerate}

\usepackage{xcolor}\colorlet{linkequation}{blue}
\usepackage[colorlinks=true, 
            linkcolor=red,   
            filecolor=red,   
            citecolor=green, 
            urlcolor=blue,   
            hyperfootnotes=true]{hyperref}
\usepackage{footnotebackref}

\newcommand*{\SavedEqref}{}
\let\SavedEqref\eqref
\renewcommand*{\eqref}[1]{%
  \begingroup
    \hypersetup{
     linkcolor=linkequation,
      linkbordercolor=linkequation,
    }%
    \SavedEqref{#1}%
  \endgroup
}

\newcommand\ba{\begin{eqnarray}}
\newcommand\ea{\end{eqnarray}}

\newcommand{\gevc}{\mbox{~GeV/c}}
\newcommand{\mb}{\mbox{~mb/GeV$^4$}}

\begin{document}

\title{The production of charged-kaon pairs in proton-antiproton collisions: The role of higher-twist mechanism}

\author{M.~Demirci}
\email{mehmetdemirci@ktu.edu.tr}
\affiliation{Department of Physics, Karadeniz Technical University, TR61080 Trabzon, Turkey}%
\author{A. I.~Ahmadov}
\email{ahmadovazar@yahoo.com}
\affiliation{Department of Theoretical Physics, Baku State University, AZ1148 Baku, Azerbaijan}%

\date{\today}

\begin{abstract}
The higher-twist (HT) contribution to the charged kaon pair production in the high energy proton-antiproton collisions at large transverse momentum $p_T$ is
investigated by using the frozen coupling constant approach for various kaon distribution amplitudes (DAs),
which are predicted by light-cone formalism, the light-front quark model, the nonlocal chiral quark model and the light-front holographic AdS/CFT approach.
In the numerics the dependencies of the HT contribution on the transverse momentum $p_T$, the rapidity $y$, and the variable $x_T$ are discussed with special emphasis put on DAs.
The HT contribution is also compared with the leading-twist ones. It is shown that the HT contributions are dependent on the kaon DAs and also some other phenomenological parameters such as momentum cut-off parameter $\Delta p$.
Inclusive kaon pair production presents a remarkable test case in which HT terms dominate those of
LT in certain kinematic regions. The HT direct production process via gluon-gluon fusion contributes significantly to the inclusive cross section at large $p_T$.
\end{abstract}

 \maketitle

\section{\bf Introduction}
The hadron production has been investigated for a long period in high-energy physics and nuclear physics,
as well as cosmic-ray physics. The absolute yields and the transverse momentum ($p_T$) spectra
of identified hadrons are among the fundamental physical observables in high-energy hadron-hadron collisions.
These observables could be used to check and refine phenomenological models of the strong interaction. Furthermore,
the search for large-$p_T$ hadron-hadron productions has contributed essentially to our understanding of the nature of
short-distance parton-parton interactions. Particularly with the
advent of the high-energy proton-antiproton collisions, such
interactions have been successfully explained by using the well-known techniques of perturbation theory.

In the standard perturbative Quantum Chromodynamics (pQCD) picture, hadrons are produced by the parton jet fragmentation.
However, higher-twist (HT) processes can also be used as production mechanism.
The term ``twist'' emerged in the operator product expansion (OPE), which was a method used for obtaining
predictions of pQCD in deep inelastic scattering~\cite{Owens}.
Today, the term refers to contributions suppressed by powers of large momentum with respect to the leading terms.
The leading-twist (LT) is standard processes of the pQCD within the collinear factorization,
where hadrons are produced through fragmentation processes. On the other hand, HT processes are taken usually
as direct hadron production, in which the hadron is produced directly in the hard
subprocess rather than by quark/gluon fragmentation~\cite{Brodsky2010}.

In the last forty years, HT effects in QCD have been investigated by many researchers
for various phenomena (see, e.g., Refs.~\cite{Baier,Berger,Bagger,Bagger1,Pohjoisaho,Ahmadov}).
The results of these studies show that, the HT contributions to the cross sections and
other characteristics of different processes may be considerable in some regions of the phase space,
and the HT contributions are strongly dependent on the
choice of the hadronic wave functions, hence the distribution amplitudes (DAs). The hadronic DAs in view of internal structure
degrees of freedoms are essential for obtaining accurate predictions in QCD. The HT processes have also importance in understanding of “Baryon anomaly”,
appeared in measurements of large-$p_T$ hadron production at RHIC~\cite{Brodsky2008}. More research is needed to clarify the nature of the  HT effects in QCD. Meson pair production in a hadron collider at large-$p_T$ can be used as a short distance probe of the incident hadrons.

In the present work, we examine the HT effect on charged-kaon pair production at proton-antiproton collisions for different kaon
DAs predicted by pQCD evaluation, light-cone formalism, the light-front quark model, the nonlocal chiral quark model
and the light-front holographic AdS/CFT correspondence. The physical information of the inclusive kaon pair production can be obtained
efficiently in the pQCD and it is, hence, possible to compare directly with the experimental data.
The corresponding hard-scattering subprocesses occur via three different mechanisms:
Direct production (kaons are produced directly at the hard-scattering subprocess),  semi-direct production (one kaon is produced from jet fragmentation, while the other one is directly produced)
and double jet production and fragmentation (both kaons are produced from fragmentation of the final quarks or gluons).
The first two mechanisms are of HT,
while the last one corresponds to LT contributions. Therefore, we must systematically compare these different mechanisms.

We use the frozen coupling constant (FCC) approach during numerical evaluation. Although the FCC approach was introduced a long time
ago~\cite{Curci1,Curci2,PLUTO}, it is still interesting in nowadays~\cite{Webber,Ciafaloni,
Kotikov,Ermolaev}. For first time, it has originated from the divergent infrared behavior of the
renormalization group expression for $\alpha_{s}$. The FCC can be used in the
infrared domain since it is a constant. The other reason for using this approach
is that the pQCD coupling is running, and the effects of running $\alpha_{s}$
should be taken into account in every calculation. On the other hand, this makes
some QCD calculations very difficult. However, for approximate predictions, it may be convenient to use some effective
coupling which imitates the running of $\alpha_{s}$ in the perturbative domain. To get an agreement with experimental data,
the value of the FCC is generally set from purely phenomenological predictions. Furthermore, it is used in combination with other phenomenological parameters to define hadronic processes.
The fixed $\alpha_{s}$ has been used in various calculations carried out in the
framework of the leading logarithmic approximation where the most important logarithmic contributions are completely resummed
whereas argument of $\alpha_{s}$ is set off a posteriori from physical predictions.

The another way is through solution of the Schwinger-Dyson equations (SDE) for investigating the infrared behavior of
the running coupling constant, gluon (and ghost) propagator at low energies~\cite{Roberts}.
In order to get infrared finite propagators, one can use a method where the gluon acquires a dynamical mass $m_{\text{g}}^2$ (see, e.g., Ref.~\cite{Cornwall}), and the another is that the gluon propagator goes to zero when the momentum
$Q^2\rightarrow 0$ (discussed in Refs.~\cite{Alkofer, Smekal}). In both cases,
there appear the freezing of coupling constant in the infrared domain. In the case where squared momentum of
hard gluon gets the form $Q^2 \to Q^2+m_{\text{g}}^2$, argument of running coupling constant takes also the same form. Here
$m_{\text{g}}$ is interpreted as an effective dynamical mass of gluon. 

The experimental researches on measurements of charged hadrons (or charged tracks) in proton-antiproton collisions
have been carried out at $\sqrt{s}= 0.630, 1.8, 1.96$ TeV by CDF~\cite{CDF1,CDF2,CDF1960} and
$\sqrt{s}= 0.5, 0.9, 7$  TeV by UA (CMS)~\cite{UA1}. For different center of mass energies,
the differential cross sections are constructed and
compared to a scaling with the variable $x_T=2 p_T/\sqrt{s}$.
We provide our calculations at $\sqrt{s}=500$ GeV.
To compared with other energies we also present distribution of the variable $x_T$ for a given $p_T$.

Kaon pair production in photon-photon and proton-antiproton collisions have been studied
from high to low energies during the last years, using different methods such as  HT mechanism, central exclusive production mechanism, effective meson theory, and standard pQCD (see,~\cite{aaTOMM,Bystritskiy1,Khoze,Djagouri} and references therein).

The present work is organized as follows.  In the next section, we provide some expressions for the HT (in Sec.~\ref{sec:ht}) and
LT (in Sec.~\ref{sec:lt}) contributions to cross section of the process $p\bar {p}\to K^{+}K^{-}X$ and a brief review
for kinematics variables and convolution of contributions (in Sec.~\ref{sec:tcs}). In Sec.~\ref{sec:DA}, we give some DAs of kaon used in our calculations and their evolutions according to the scale $Q$. In Sec.~\ref{sec:results},
we present numerical results and discuss the dependence of the cross sections on the kaon DAs and other physical parameters in detail.
Finally, the summary and concluding remarks are given in Sec.~\ref{sec:conc}.

\section{The Analytical Results of Kaon Pair Production}\label{sec:cs}
The almost scale-invariant behavior of two-particle (gluon and quark) hard-scattering processes
is a fundamental property of asymptotic freedom and QCD. If these hard-scattering subprocesses are convoluted
with the parton distribution functions (PDFs) of the initial hadrons and
the fragmentation functions (FFs) which produce final state interactions, the resulting inclusive cross section scales as $1/p_T^n$.

In the present study, we aim to investigate the inclusive production of kaon pairs with large-$p_T$ in $p\bar{p}$ collisions. For this, we consider HT contributions to the cross section
by using the FCC approach for different kaon DAs. Furthermore, HT contributions are compared with LT ones.
This comparison will allow us to determine such regions in the phase space
where HT contributions are essentially observable.
In order to obtain an accurate value of the ratio contributions of HT and LT,
we use the fact that prompt kaons are ``non-accompanied" by any other hadron.
However, this is not valid for the general case in which particles are occurring from the jets fragmentation.
This criterion can be incorporated into the general formulas via a momentum cut-off parameter $\Delta p$ \cite{Engels}.

We present details of analytical calculations on HT and LT contributions in the following subsections.

\subsection{Higher-Twist Contributions}\label{sec:ht}
\begin{figure}[!t]
    \begin{center}
\includegraphics[scale=0.63]{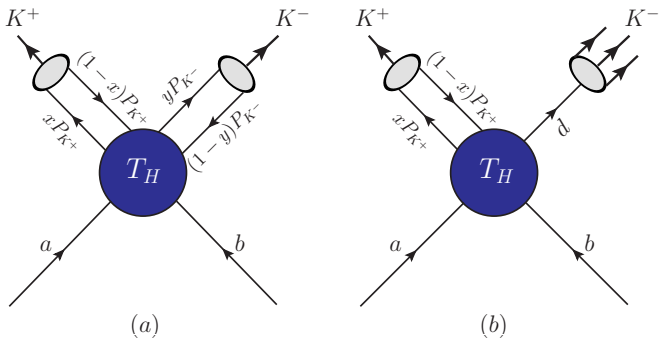}
 \end{center}
\caption{Factorization of the kaon pair production amplitude in QCD at large momentum transfer
in case of (a)\label{fig:fig1a} direct production and (b)\label{fig:fig1b} semi-direct production. The labels $a, b, d$ represent quarks or gluons.}
\end{figure}

For HT contribution to the charged kaon pair production, there are two different mechanisms included in the hard scattering subprocess:
\begin{itemize}
  \item[\textit{i.}] \textit{Direct-production} (Fig.~\ref{fig:fig1a}a): Both kaons are produced directly,
  \item[\textit{ii.}] \textit{Semi-direct production} (Fig.~\ref{fig:fig1b}b): One kaon is produced directly and the other one is produced from jet fragmentations.
\end{itemize}
The differential cross section for partonic subprocess is given by
\ba
\frac{d\hat{\sigma}(a b \to c d)}{d\cos\theta}=\frac{1}{32 \pi \hat{s}} \sum|\overline{\mathcal{M}}|^2
\ea
where $\mathcal{M}$ is the invariant amplitude of the HT hard-scattering subprocesses and
the bar on it refers to the average over initial spins and colours.

In order to obtain the corresponding amplitude $\mathcal{M}$,
one should take integrations with the kaon DAs over
the longitudinal momentum fractions $x_i$ and $y_i$ carried by the kaons's quark and antiquark.
In light of this discussion, it takes the following form~\cite{Lepage2}:
\ba
\begin{split}
\mathcal{M}(\hat{s},\theta)=&\int_{0}^{1}{[dx_i]} \int_{0}^{1}[dy_i]~\Phi_{K^{-}}(y_i,\widetilde{Q}_y)
\\
&\times  T_{H}(x_i,y_i;\hat{s},\theta)~\Phi_{K^{+}}(x_i,\widetilde{Q}_x)
\end{split}
\ea
where $[dx_i]=\delta(1-\sum_{k=1}^n x_k)\prod_{k=1}^n dx_k$ and $n$ is the number of the valance quarks.
The scale $\widetilde{Q}$ can be taken as $\widetilde{Q}_x=min(x,1-x)Q$ and similarly $\widetilde{Q}_y=min(y,1-y)Q$.
$T_H$ is the hard-scattering amplitude of subprocess for the production of the valance quarks collinear with each kaon. $\Phi(x,\widetilde{Q}_x)$ is the quark distribution amplitude of the kaon (see Sec.~\ref{sec:DA} for details),
sharing fractions $x$ and $(1-x)$ of the kaon's total momentum.
Similarly, $\Phi(y,\widetilde{Q}_y)$ is sharing fractions $y$ and $(1-y)$ of the other kaon's total momentum.
They were integrated over transverse momenta $k_T<Q$.

In pQCD calculations, the amplitude $T_{H}$ at the leading order strongly depends on the
renormalization scale, but does not depend
on the factorization scale. However, one-loop QCD corrections to $T_{H}$ lead to its explicit dependence on the both scales.
Moreover, it should be noted that both scales
can be chosen autonomously since they are independent of each other.
In principle, under any choice of renormalization scheme and scale, all measurable quantities in QCD must be invariant.
The use of different schemes and scales can lead to different
theoretical predictions.  Therefore, the constructive mathematical
apparatus for defining QCD is a choice of the renormalization scale
which makes scheme independent results at all fixed order in running
coupling constant $\alpha_{s}$.
\begin{figure*}[!htb]
    \begin{center}
\includegraphics[scale=0.55]{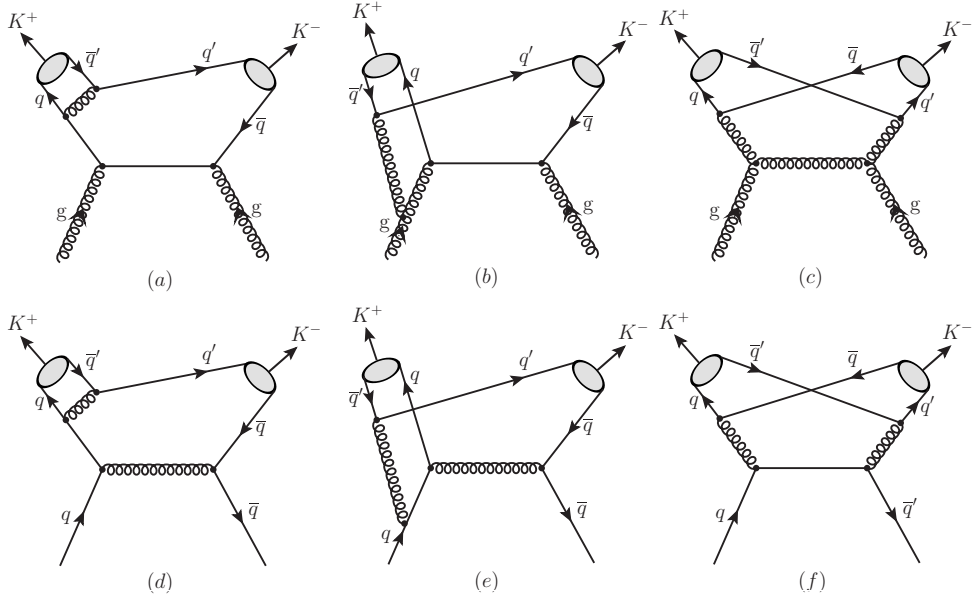}
     \end{center}
\caption{QCD Feynman diagrams  of the subprocesses
$\text{g}\text{g} \to K^+K^-$ and $q\bar q \to K^+K^-$ for
direct kaon pair production at leading order. The gray ovals indicate to the wave functions of kaons.}\label{fig:fig2}
\end{figure*}

Let us now give more details on the hard-scattering subprocesses for direct and semi-direct productions of the
charged K-meson pair. To make the analysis of collinear divergences easier,  we assume that both kaons are emitted
at $\cos(\theta)=0$ ($\theta$ is the emission angle measured in the center-of-mass frame), with equal $p_T$.
We neglect all quark and kaon masses in all diagrams contributing to the hard-scattering subprocess at leading order,
resulting in errors only of order $m^2 /s \ll 1$.

\textit{i. For direct kaon pair production,  we take the following hard-scattering subprocesses:}
\begin{itemize}
  \item[$\diamond$] $\text{g}\text{g}\to K^+K^-$,
  \item[$\diamond$] $q\bar{q}\to K^+K^-$ for $q=u$ and $s$.
\end{itemize}
We show some Feynman diagrams for these subprocesses in Fig.~\ref{fig:fig2}.
At each vertex (where three lines join) the interaction is proportional to the QCD coupling constant $\alpha_s$,
so if the cross section of the direct-production process is computed,
it would be ended up with a number proportional to the 4 power of $\alpha_s$.
For direct kaon pair production, after averaging over
colors and spins of incoming particles, the associated
differential cross sections are written via the electromagnetic form factor of the kaon $F_K$ as follows:
\begin{widetext}
\begin{equation} \label{eq:ggKK}
\begin{split}
\frac{d\hat{\sigma}(\text{g}\text{g}\to K^+K^-) }{d\cos\theta}=
\frac{ \pi \alpha_{s}^2 F_{K}^2}{18 \hat{s}} \biggl[\frac{1}{I_{K}}\int_{0}^{1}dx\int_{0}^{1}dy \frac{\Phi_{K}(x,\widetilde{Q}_x)\Phi_{K}(y,\widetilde{Q}_y)}{x(1-x)y(1-y)}
\frac{x(1-x)+y(1-y)}{x y+(1-x)(1-y)}\biggr]^2,
\end{split}
\end{equation}

\begin{equation} \label{eq:qqKK}
\begin{split}
\frac{d\hat{\sigma}(q\overline q \to K^+K^-) }{d\cos\theta}=
\frac{\pi \alpha_{s}^2 F_{K}^2}{972 \hat{s}}  \biggl[\frac{1}{I_{K}}\int_{0}^{1}dx\int_{0}^{1}dy &\frac{\Phi_{K}(x,\widetilde{Q}_x)\Phi_{K}(y,\widetilde{Q}_y)}{x(1-x)y(1-y)}\biggl(7-16x y\\
&-\frac{2x (1-2y(x+y))-4x^2+4x y}{x y+(1-x)(1-y)}\biggr)\biggr]^2.
\end{split}
\end{equation}
\end{widetext}
where $F_{K}$ and $I_{K}$ are given by
\ba
F_{K}(\hat{ s})=\frac{16\pi \alpha_{s}}{3\hat{ s}} \frac{f_{K}^2}{12} I_K^2,\\
I_{K}=\int_{0}^{1} \frac{\Phi_{K}(x,\widetilde{Q}_x)}{x(1-x)}dx.
\ea
The leading order hard scattering amplitudes exhibit divergence at both end points of $x$ and $y$.
However, the end point singularities are suppressed by the bound state quark DAs.

\begin{figure*}[!htb]
    \begin{center}
\includegraphics[scale=0.55]{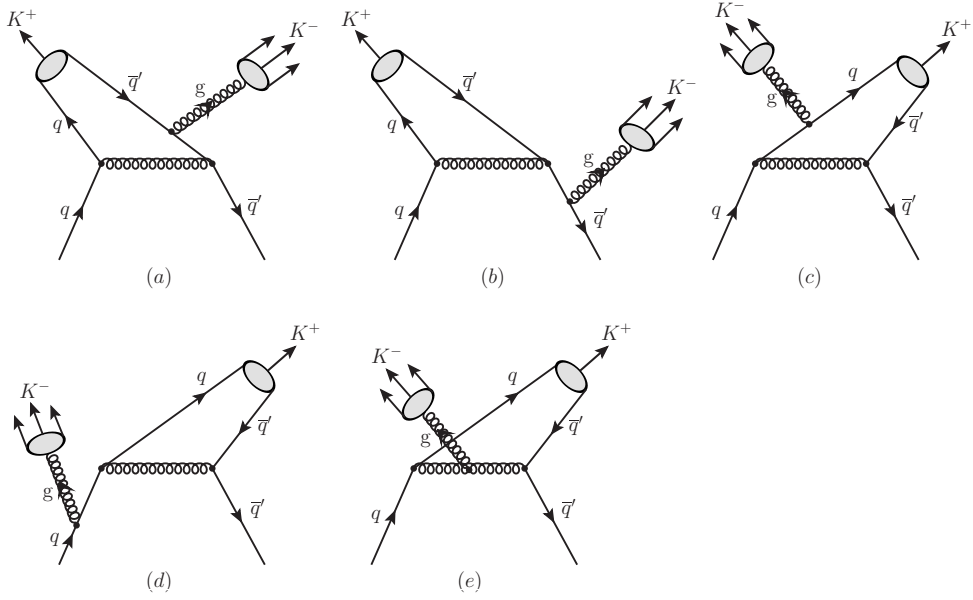}
     \end{center}
\caption{QCD Feynman diagrams of the subprocess $q\bar q' \to
K^+\text{g}$ (where the gluon is fragmented to kaon $K^-$) for semi-direct kaon pair production at leading level. The gray ovals indicate to the corresponding wave functions.
Additionally, the diagrams corresponding to the subprocess $q\text{g}\to K^{+}q'$
(where the quark $q'$ is fragmented to kaon $K^-$) can be plotted by exchanging the outgoing gluon and the incoming antiquark lines (which become a quark line) in each of above diagrams. }\label{fig:fig3}
\end{figure*}

\textit{ii. There are two types of subprocesses for the semi-direct production:}
\begin{itemize}
  \item[$\diamond$] $q\bar{q}' \to K^\pm \text{g}$ where the gluon is fragmented to Kaon $(\text{g}\Rightarrow K^\mp)$,
  \item[$\diamond$] $q\text{g}\to K^{\pm}q'$ (and $\bar{q} \text{g}\to K^{\pm}\bar{q}'$) where the final quark is fragmented to Kaon $(^(\bar{q}'^) \Rightarrow K^{\mp})$.
\end{itemize}
In Fig.~\ref{fig:fig3}, we show some Feynman diagrams for subprocesses of semi-direct production.
We note that there are crossing symmetry among the invariant amplitudes associated with the subprocesses $q\bar{q}' \to K^\pm \text{g}$ and $q\text{g}\to K^{\pm}q'$, which can
be directly checked by crossing exchanges $\hat{s} \leftrightarrow -\hat{t}$ at fixed $\hat{u}$ in the invariant amplitude squared (summed over spin and color indices).
At each vertex the interaction is proportional to the coupling constant $\alpha_s$, so the cross section of semi-direct production process would be end up with a number proportional to the 3 power of $\alpha_s$ as seen in Fig.~\ref{fig:fig3}.
Summing overall diagrams of either type, the corresponding differential cross sections of semi-direct production for each subprocess are given by
\begin{widetext}
\ba
\begin{split} \frac{d\hat{\sigma}(q\overline{q}'\to K^\pm
\text{g})}{d\cos\theta}&=\frac{32 \pi \alpha_{s}^2}{81{\hat
s}} \frac{16\pi \alpha_{s}}{3\hat{ s}} \frac{f_{K}^2}{12} \biggl[\int_{0}^{1} dx \frac{\Phi_{K}(x,\widetilde{Q}_x)}{x(1-x)}\biggr]^2,
\end{split}
\ea
\ba \frac{d\hat{\sigma}(q\text{g}\to K^\pm
q')}{d\cos\theta}=\frac{5 \pi \alpha_{s}^2}{108{\hat
s}}\frac{16\pi \alpha_{s}}{3\hat{ s}} \frac{f_{K}^2}{12}  \biggl[\int_{0}^{1} dx \frac{\Phi_{K}(x,\widetilde{Q}_x)}{x(1-x)}\biggr]^2,
\ea
\ba \frac{d\hat{\sigma}(\overline{q} g\to K^\pm \overline
q')}{d\cos\theta}=\frac{5 \pi \alpha_{s}^2}{108{\hat
s}} \frac{16\pi \alpha_{s}}{3\hat{ s}} \frac{f_{K}^2}{12} \biggl[\int_{0}^{1} dx \frac{\Phi_{K}(x,\widetilde{Q}_x)}{x(1-x)}\biggr]^2.
\ea
\end{widetext}
The initial $q, \bar{q}$ and g are the constituent of the initial target proton and
anti-proton, respectively.

\subsection{Leading-Twist Contributions}\label{sec:lt}
It is also essential to examine effects of the HT contributions as well as to compare of HT contributions with LT ones for problems of the pQCD.
\begin{figure}[!hbt]
    \begin{center}
\includegraphics[scale=0.65]{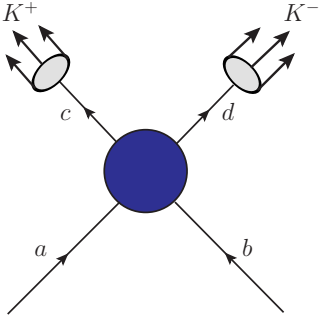}
     \end{center}
\caption{\label{fig:fig4} A general diagram for the leading-twist
subprocesses $ab \to cd$ where the final partons are fragmented into the charged kaon pairs, separately.}
\end{figure}
From this comparison we can determine such regions in the phase space
where HT contributions are actually observable. For LT contributions to
the charged \textit{K}-meson pair production in $p\bar{p}$ collisions, we consider the following hard subprocesses:
\begin{enumerate}
  \item[$\diamond$] $q\bar{q}\to q\bar{q}$     ~: $q \Rightarrow K^+$, $\bar q \Rightarrow K^-$,
  \item[$\diamond$] $q\bar{q}\to q'\bar{q}'$    : $q'\Rightarrow K^+$, $\bar{q}' \Rightarrow K^-$ ,
  \item[$\diamond$] $q\bar{q}'\to q\bar{q}'$    : $q\Rightarrow K^+$, $\bar{q}' \Rightarrow K^-$ ,
  \item[$\diamond$] $q\bar{q}\to \text{g}\text{g}$            : $\text{g}\Rightarrow K^+$, $\text{g} \Rightarrow K^-$,
  \item[$\diamond$] $q\text{g}\to q\text{g}$                  : $q\Rightarrow K^+$, $\text{g}\Rightarrow K^-$,
  \item[$\diamond$] $\text{g}\text{g} \to \text{g}\text{g}$                 : $\text{g}\Rightarrow K^+$, $\text{g} \Rightarrow K^-$ and
  \item[$\diamond$] $\text{g}\text{g} \to q\bar{q}$           : $q\Rightarrow K^+$, $\bar q \Rightarrow K^-$,
\end{enumerate}
where the symbol ``$\Rightarrow$" represents fragmentation. We plot a representative diagram for these subprocesses in Fig.~\ref{fig:fig4}.
In Table~\ref{tab:table1}, we list the associated expressions for differential cross sections of the LT subprocesses given in~\cite{Owens}.
The initial and final state  colors and spins have been averaged
and summed, respectively. The cross sections for these QCD-hard subprocesses are dominated by $\hat{t}$-channel gluon exchange contributions.
\begin{table}[ht]
\caption{The associated differential cross sections for the leading-twist subprocesses.
The primed symbol ($q'$) denotes distinct flavor and $\hat{s}, \hat{t}$, $\hat{u}$ are the Mandelstam variables of the subprocess.}\label{tab:table1}
\begin{ruledtabular}
\begin{tabular}{ll}
\multicolumn{1}{c}{$ab \to cd$}&  \multicolumn{1}{c}{$\frac{d\hat{\sigma}(ab \to cd)}{d\cos\theta}$}  \\\hline
$q\bar{q} \to q\overline{q}$& $\frac{\pi\alpha_{s}^2}{2{\hat
s}}\frac{4}{9} \biggl(\frac{\hat{u}^2+\hat{s}^2}{\hat{t}^2}+\frac{\hat{u}^2+\hat{t}^2}{\hat{s}^2}-\frac{2}{3} \frac{\hat{u}^2}{\hat{s}\hat{t}}\biggr)$ \\
$q \bar{q} \to q'\bar{q}' $& $\frac{2\pi\alpha_{s}^2}{{9\hat s}}\left(\frac{\hat{u}^2+\hat{t}^2}{\hat{s}^2}\right)$   \\
$q \overline{q}' \to q \overline{q}'$&$\frac{2\pi\alpha_{s}^2}{{9\hat s}}\left(\frac{\hat{u}^2+\hat{s}^2}{\hat{t}^2}\right)$ \\
$q\bar{q} \to \text{g}\text{g}$ & $\frac{\pi\alpha_{s}^2}{2{\hat
s}}\frac{8}{3}\left(\frac{4}{9} \frac{\hat{u}^2+\hat{t}^2}{\hat{u}\hat{t}}- \frac{\hat{u}^2+\hat{t}^2}{\hat{s}^2}\right)$ \\
$q\text{g}\to q\text{g}$& $\frac{\pi\alpha_{s}^2}{2{\hat
s}}\left(\frac{\hat{u}^2+\hat{s}^2}{\hat{t}^2}-\frac{4}{9} \frac{\hat{u}^2+\hat{s}^2}{\hat{u}\hat{s}}\right)$ \\
$\text{g}\text{g}\to \text{g}\text{g}$ &$\frac{\pi\alpha_{s}^2}{{\hat s}}\frac{9}{4}\left(3-\frac{\hat{u}\hat{t}}{\hat{s}^2}-\frac{\hat{u} \hat{s}}{\hat{t}^2}-\frac{\hat{s}\hat{t}}{\hat{u}^2}\right)$\\
$\text{g}\text{g} \to
q\bar{q}$&$\frac{\pi\alpha_{s}^2}{2{\hat s}}\left(\frac{1}{6} \frac{\hat{u}^2+\hat{t}^2}{\hat{u}\hat{t}}-\frac{3}{8} \frac{\hat{u}^2+\hat{t}^2}{\hat{s}^2}\right)$\\
\end{tabular}
\end{ruledtabular}
\end{table}
If the kaons are produced at emission angle $\theta=90^\circ$ with
the rapidities of final particles $y_1=y_2=0$, the hard scattering cross section
$d\sigma/d\hat t$ is actually probed at angles around $\theta=90^\circ$, hence
$\hat t=\hat u =-\hat s/2$.

The LT contributions to production of kaon pairs at large $p_T$ in proton-antiproton collisions are conventionally analyzed
within the framework of pQCD by convoluting the hard subprocess cross sections given in Table~\ref{tab:table1} with evolved
FFs and PDFs.

\subsection{The Convolution of Twist Contributions in Proton-Antiproton Collisions}\label{sec:tcs}
Let us now consider the hadronic process of the charged K-meson pair production
\ba
p \bar p \to K^{+}K^{-}+X \label{eq:hprocess}
\ea
where $X$ indicates all other particles in the final state. We assume that both kaons in the $p\bar{p}$ collisions are emitted at $90^\circ$ in
the center-of-mass frame, with equal $p_T$. We apply the factorization formula predicted by Gunion and Petersson~\cite{Petersson}.
In this approach, in order to obtain inclusive production of the charged-kaon pair~\eqref{eq:hprocess},
differential cross section of the corresponding hard-scattering subprocess\footnote{We indicate the higher-twist cross
section by $ \Sigma_{K^+K^-}^{HT} $ and the leading-twist cross
section by $ \Sigma_{K^+K^-}^{LT} $.} is convoluted with the two PDFs and two FFs:
\begin{widetext}
\begin{equation} \label{eq:ICS}
\begin{split}
\Sigma_{K^{+}K^{-}}\equiv\frac{EE' d\sigma (p \bar p \to K^{+}K^{-}
X)}{d^3pd^3p'} = \frac{1}{(\pi \sqrt{s})^2 <k^{2}_T>}&\int_{z_{min}}^{1}\frac{dz}{z^2}\int_{z_{min}}^{1}\frac{dz'}{z'^2}
F(z,z')G_{{a}/{p_{1}}}(x_{1},\mu_{F}^2) G_{{b}/{p_{2}}}(x_{2},\mu_{F}^2) \\
&\times\frac{d\hat{\sigma }(a b \to c d) }{d\cos\theta}D^{K^{+}}_{c}(z,Q^2) D^{K^{-}}_{d}(z',Q^2),
\end{split}
\end{equation}
\end{widetext}
where $\sqrt{s}$ is the center-of-mass energy of main process,
$<k^{2}_T>$ is the mean square of the intrinsic partonic momentum for partons $a,b$. The functions $G_{a/p_1}$ and $G_{b/p_2}$
are the universal PDFs for the partons $a,b$ in the
proton and antiproton $p_1$, $p_2$,
respectively. They depend on the longitudinal momentum fractions of
the two partons in case final jets fragmenting into kaon pairs,
$x_1= x_2\cong2p_T/\sqrt{zz's}$. Dynamical properties of the jets are close to the parton carried a fraction of momentum
of parent hadron.
The correlation function $F(z,z')$ is denoted by
\ba
F(z,z')=\frac{z+z'}{2\sqrt{zz'}}\exp{\left[\frac{-(z-z')^2p_{T}^2}{2z^2z'^2 <k^{2}_T>}\right]}.
\ea

In our numerical evaluations, the functions $D^{K^{+}}_{c}(z,Q^2)$ and $D^{K^{-}}_{d}(z',Q^2)$ in Eq.~\eqref{eq:ICS} are taken for each different production mechanism as follows:
\begin{itemize}
  \item $D^{K^{+}}_{K^{+}}(z,Q^2)=\delta(1-z)$ and $D^{K^{-}}_ {K^{-}}(z',Q^2)=\delta(1-z')$ in case of direct kaon pair production ($c\equiv K^+$ and $d\equiv K^-$),
  \item $D^{K^{+}}_{K^{+}}(z,Q^2)=\delta(1-z)$, whereas $D^{K^{-}}_{d}(z',Q^2)$ is the usual FF in case of semi-direct kaon pair production ($c\equiv K^+$ and $d \equiv \bar{u}, s, g$),
  \item $D^{K^{+}}_{c}(z,Q^2)$ and $D^{K^{-}}_{d}(z',Q^2)$ are the usual FFs for leading-twist contributions,
\end{itemize}
where $D^{K}_{c,d}$ indicates the quark fragmentation function into a kaon including a quark of the same flavor. For leading-twist subprocesses, the kaons are indirectly emitted from the
final partons with fractional momentums $z,z'$. In the numerical treatment, we use the usual FFs in~\cite{FFs}, parameterized as
\ba D^{h}_{c,d}(z,Q^2)=N z^\alpha (1-z)^\beta (1+z)^\gamma.
\ea
Furthermore, we use the MSTW2008 PDFs\cite{MSTW} for the quark and gluon distribution
functions inside the proton and antiproton.

The minimum value of the momentum fraction of the final
parton is defined in this form:
\ba z_{min}=\frac{p_T}{p_T+\triangle p}
\ea
where momentum cut-off parameter $\triangle p$ defines
the experimental upper limit for non-detection of one or more
particles accompanying either kaon detected. When this limit is exceeded, the corresponding event will be refused.
The prompt kaons appear non-accompanied by any other hadron,
but this is not the case, in general, for particles produced from jet fragmentation.

The longitudinal momentum fractions of partons are defined in the following
forms
\ba
x_1=\frac{p_{T}}{\sqrt{s}}(e^{y_1}+e^{y_2}),~\\
x_2=\frac{p_{T}}{\sqrt{s}}(e^{-y_1}+e^{-y_2}),
\ea
in which $y_1$, $y_2$  are the rapidities of the final particles. We use the kinematic expressions discussed in Ref.~\cite{Owens} and in our previous works~\cite{Demirci1,Demirci2,Demirci3}.

We note that the higher-twist cross sections are proportional to $\hat{s}^{-4}$ for direct-production and $\hat{s}^{-3}$
for semi-direct production, hence, they have the form of $p_T^{-8}$ and $p_T^{-6}$, respectively. However,
the $p_T^{-6}$ processes $q\bar{q} \rightarrow Mq$ and $\text{g}q \rightarrow Mq$ are interesting
in high-$p_T$ meson production processes such as $pp \rightarrow MX$ because the
meson is produced directly in the subprocess without the fragmentation. In fact the contributions of
standard $p_T^{-4}$ scaling processes such as $qq \rightarrow qq$, $\text{g}\text{g} \rightarrow \text{g}\text{g}$ and
$\text{g}q \rightarrow \text{g}q$ are highly suppressed by 2 to 3 orders of magnitude relative to the "directly coupled" contributions because of the suppression
of jet fragmentation $D^M_q(z)$ at large momentum fraction $z$ and the
fact that the subprocesses must arise at a remarkably larger
momentum transfer than that of the triggered particle~\cite{Ellis,ChapBrodsky}.

\section{Kaon Distribution Amplitudes and Their Evolutions}\label{sec:DA}

The important aspect of the present study is to select of the distribution amplitudes. DAs are intrinsically nonperturbative; they include all effects of collinear singularities, meson bound-state dynamics,
nonperturbative interactions and confinement. The DA $\Phi_{K}(x,\mu^2)$ is the amplitude for the kaon consisting
of a $q\bar{q}'$ pair, with the $q$  and $\bar{q}'$  collinear and on shell relative to the scale $\mu$.
During the past few decades, there have been many theoretical efforts to calculate the kaon DA using different methods such as lattice calculation, the QCD sum rule~\cite{Chernyak:1981jd,Chernyak:1982it,Chernyak:1983ej}, the chiral-quark model, and the light-front quark model. In this study we choose several DAs which show significant differences compared to each other as follows: the asymptotic DA derived in pQCD evaluation~\cite{Lepage1}, kaon DAs with
six non-trivial Gegenbauer coefficients $a_1$, $a_2$, $a_3$, $a_4$,
$a_5$, and $a_6$ derived by using the light-cone formalism (LCQM)~\cite{LCQM},
obtained from the Gaussian wave function with harmonic oscillator potential and power-law wave function (HOP and PL, respectively)~\cite{LPHOP},
and predicted within the framework of the nonlocal chiral quark model from the instanton vacuum ($\chi$QM)~\cite{XQM}.

The asymptotic DA is given by
\ba \Phi_{asy}^{K}(x)=\sqrt{3}f_{K}x(1-x), \label{eq:asy}
\ea
where $f_{K}=156.01$ MeV. The overall normalization is set by the kaon decay constant via
\ba \label{eq:normalize}
\int_{0}^{1}\Phi_{K}(x,\mu^2)dx=\frac{f_{K}}{2\sqrt{3}}.
\ea
The evolution of the DA on the scale $Q^2$ is obtained by solving a Bethe-Salpeter type equation.
The most general solution is an expansion in the Gegenbauer polynomials $C_{n}^{3/2}$ as follow \cite{Lepage1,Efremov}:
\ba
\begin{split} \label{eq:GDA}
\Phi_{K}(x,Q^2)=&\Phi_{asy}^{K}(x)
\biggl[1+\sum_{n=1}^{\infty}a_{n}(Q^2) \\
&\times C_{n}^{3/2}(2x-1)\biggr]
\end{split}
\ea
where the Gegenbauer coefficients  $a_n$ (also called Gegenbauer moments) can be determined by means of Gegenbauer polynomials orthogonality condition
\ba \label{eq:ort}
\int^1_{-1}(1-\zeta^2) C_{n}^{3/2}(\zeta)C_{n'}^{3/2}(\zeta)d\zeta=\frac{\Gamma(n+3)\delta_{nn'}}{n!(n+3/2)}.
\ea
The Gegenbauer moments $a_n$ are very useful in investigating of the DAs since they form the shape of the corresponding hadron wave function. It can be derived from theoretical models or extracted from the experiments. In principle, these moments show
how much the DAs deviate from the asymptotic one.

In Table~\ref{tab:table2}, we list values of the Gegenbauer moments used in our study.
The DAs are created by opening up to the first six term of Eq.~\eqref{eq:GDA} as seen in Eq.~\eqref{eq:GDA2} and
then using the values of Gegenbauer moments in the following table.
\begin{table}[ht]
\caption{The Gegenbauer moments of the HT kaon DAs obtained from the different methods at the scale $ \mu_0\sim$1 GeV.}\label{tab:table2}
\begin{ruledtabular}
\begin{tabular}{cR[.][.]{1}{2}R[.][.]{1}{4}R[.][.]{1}{4}R[.][.]{1}{5}}
Moments $a_n$ &\multicolumn{1}{c}{LCQM~\cite{LCQM}} & \multicolumn{1}{c}{HOP~\cite{LPHOP}} & \multicolumn{1}{c}{PL~\cite{LPHOP}}& \multicolumn{1}{c}{$\chi$QM~\cite{XQM}}  \\\hline
$a_1$&0.08      &-0.1501  &-0.0218   &-0.00474  \\
$a_2$&0         &-0.1474  &-0.0385   &-0.11797 \\
$a_3$&0.03       &0.0198   &-0.0003   &-0.00298  \\
$a_4$&-0.06     &-0.0162  &-0.0090   &-0.01314  \\
$a_5$&-0.14     &0.0137   &0.0004    &-0.00068  \\
$a_6$&-0.03     &-0.00036 &-0.0030   &-0.00282  \\
\end{tabular}
\end{ruledtabular}
\end{table}
Then, we label them as follows: $\Phi^{LCQM}_{K}(x,Q^2)$, $\Phi^{HOP}_{K}(x,Q^2)$, $\Phi^{PL}_{K}(x,Q^2)$, and $\Phi^{\chi QM}_{K}(x,Q^2)$.
The kaon DA can be written by the Gegenbauer-polynomial
expansion up to the sixth moment as follows:
\begin{widetext}
\ba
\begin{split} \label{eq:GDA2}
\Phi_{K}(x,\mu_0^2)&=\Phi_{asy}^{K}(x)\biggl[1+3 a_1 (2 x-1)+a_2 \left(\frac{15}{2} (2 x-1)^2-\frac{3}{2}\right)+a_3 \left(\frac{35}{2} (2 x-1)^3-\frac{15}{2} (2 x-1)\right)\\
&+a_4 \left(\frac{315}{8} (2 x-1)^4-\frac{105}{4} (2 x-1)^2+\frac{15}{8}\right)+a_5 \left(\frac{693}{8} (2 x-1)^5-\frac{315}{4} (2 x-1)^3+\frac{105}{8} (2 x-1)\right)\\
&+a_6 \frac{35}{16}\left(\frac{429}{5} (2 x-1)^6-99 (2 x-1)^4+27 (2 x-1)^2-1\right)\biggr]
\end{split}
\ea
\end{widetext}

We also use the following kaon DAs
\ba \label{eq:hol}
\Phi^{HOL}(x)=\frac{4}{\sqrt{3}\pi}\sqrt{x(1-x)}
\ea
and
\ba \label{eq:AdsCFT}
\Phi^{AdS/CFT}(x)=\frac{A_1 \kappa_1}{2\pi}\sqrt{x (1-x)}~e^{\left(-\frac{m^2 }{2\kappa_1^2 x (1-x)}\right)}
\ea
derived by light-front holographic AdS/CFT correspondence (suggested by Brodsky and T\'{e}ramond)~\cite{ADS,ADS2}.
We checked that all the DAs can be normalized by using Eq.~\eqref{eq:normalize}.
We will dwell on the dependence of cross sections upon the model structure of the above DAs.

\begin{figure}[!hbt]
    \begin{center}
\includegraphics[scale=0.40]{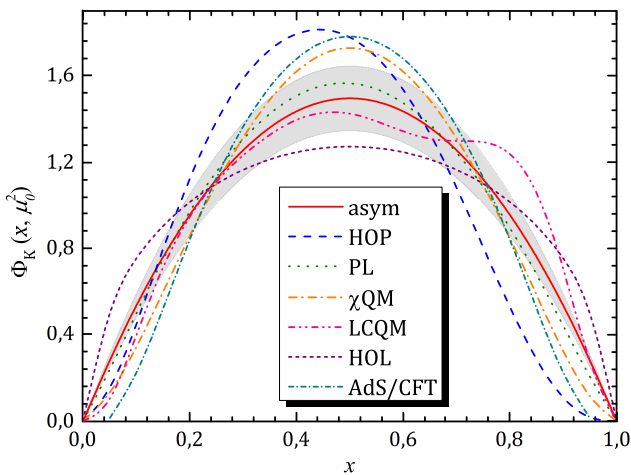}
     \end{center}
\caption{Normalized DAs for kaon obtained
from LCQM, HOP, PL, $\chi$QM, HOL and AdS/CFT models
compared with the asymptotic one (solid line). The gray band indicates $\pm10\%$ of the asymptotic DA.}\label{fig:fig5}
\end{figure}

We depict in Fig.~\ref{fig:fig5} the normalized kaon DAs for each models, using the normalization condition in Eq.~\eqref{eq:normalize}.
This figure indicate how much the DAs deviate from
the asymptotic one. The kaon DAs vanish at the endpoints ($x=0$ and $x=1$) for all cases as expected. The asymptotic DA is symmetric around $x = 0.5$.

The evolution of the DA on the factorization scale $Q^2$ is governed by the functions $a_n(Q^2)$:
\begin {equation}
a_n(Q^2)=a_n(\mu_{0}^2)\left[\frac{\alpha_{s}(Q^2)}{\alpha_{s}(\mu_{0}^2)}\right]^{\gamma_n/\beta_0},
\end{equation}
where  $\{\gamma_n\}$ are the one-loop anomalous dimensions defined by the
expression,
\ba
\gamma_n=C_F\left[1-\frac{2}{(n+1)(n+2)}+4\sum_{j=2}^{n+1}
\frac{1}{j}\right],
\ea
and $\beta_0=(11-\frac{2}{3} n_f)$ is the one-loop coefficient of QCD beta function, $n_f$ is the number of active flavors.

We note that at the boundary of $Q^2 =\mu_0^2$, Eq.~\eqref{eq:GDA} reduces to Eq.~\eqref{eq:GDA2}
with Gegenbauer moments given in Table~\ref{tab:table2}, and in the limit $Q^2\rightarrow\infty$,
Eq.~\eqref{eq:GDA} evolves into the form of the asymptotic DA~\eqref{eq:asy}, as expected.
However, with increasing $Q^2$, the evolution of DA is very slow logarithmically and, at the present-day energies, DA might be different in form.

The QCD running coupling constant $\alpha_{s}(Q^2)$ at the one-loop approximation is given as
\ba
\alpha_{s}(Q^2)=\frac{4\pi}{(11-\frac{2}{3} n_f) \ln(\frac{Q^2}{\Lambda^2})}
\ea
where $\Lambda$ is the QCD scale parameter.
The choice of renormalization scale in $\alpha_{s}(Q^2)$ is one of the main problems in QCD.
In order to make the perturbation theory meaningful, the argument of the running coupling
constant $\alpha_{s}(Q^2)$ should be fixed as the square of the momentum transfer of the exchanged gluon ~\cite{Brodsky1}.

In the expression  $\tilde{Q}=min(x,1-x)Q$,  we freeze the variable $x$ by taking its average value, namely, $\overline x =1/2$.
Additionally, the scale $Q^2$ can be taken as the average squared momentum
transfer carried by the hard gluon in a given subprocess. Within FCC approach,
we consider as follows:
 \[
\tilde{Q}=\left\{
  \begin{array}{cl}
    \frac{p_T}{2}  , & \hbox{for direct HT contribution} \\
    \frac{1}{2}\frac{p_T}{\sqrt{z}}, & \hbox{for semi-direct HT contribution} \\
    \frac{1}{2}\frac{p_T}{\sqrt{z z'}}, & \hbox{for LT contribution.}
  \end{array}
\right.
  \]

\section{Numerical Results And Discussion} \label{sec:results}
In this section, we discuss the numerical predictions for HT and LT contributions to cross section of the process  $p\bar {p}\to K^{+}K^{-}X$ in detail.
\begin{figure*}[!hbt]
    \begin{center}
\includegraphics[scale=0.40]{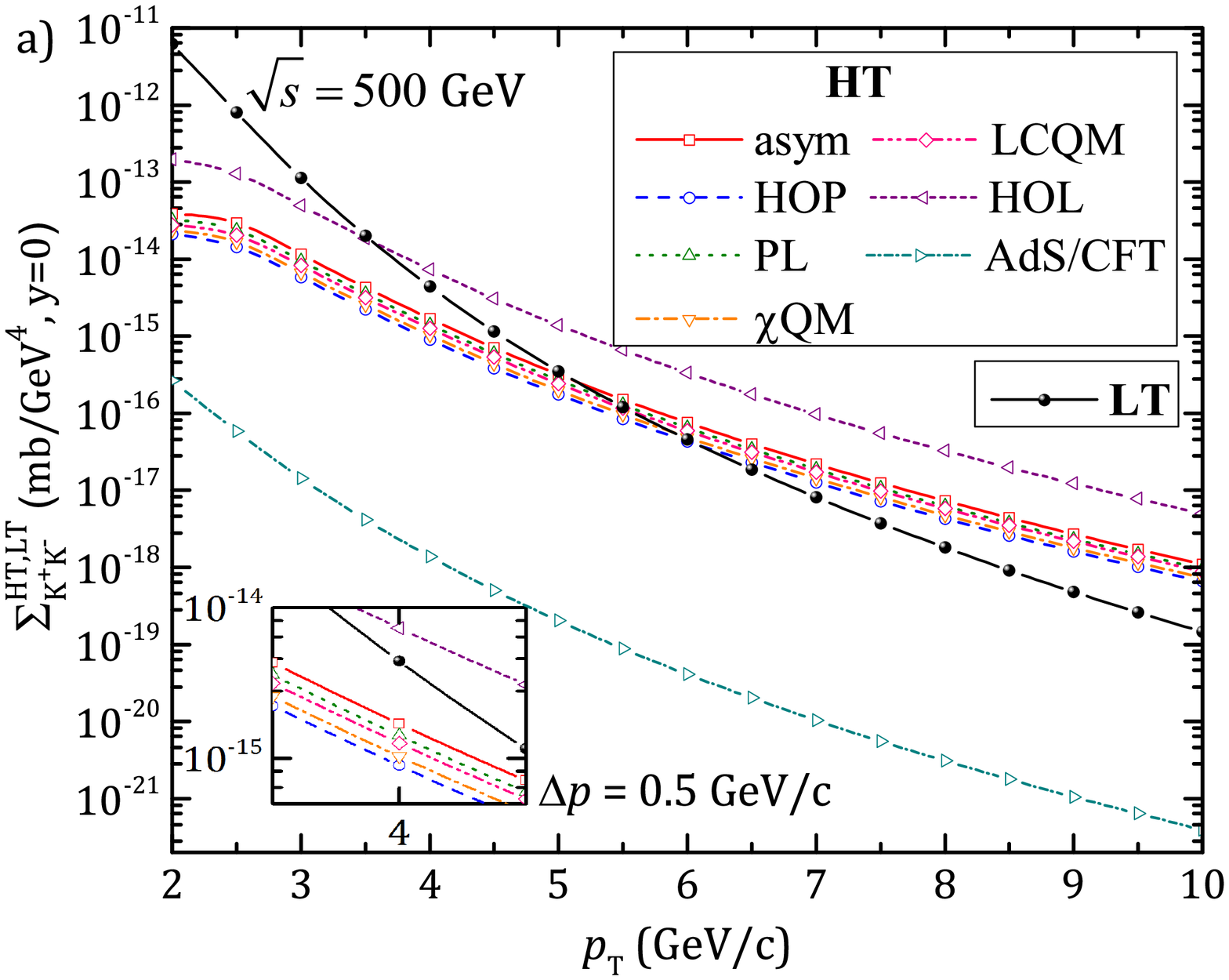}
\includegraphics[scale=0.40]{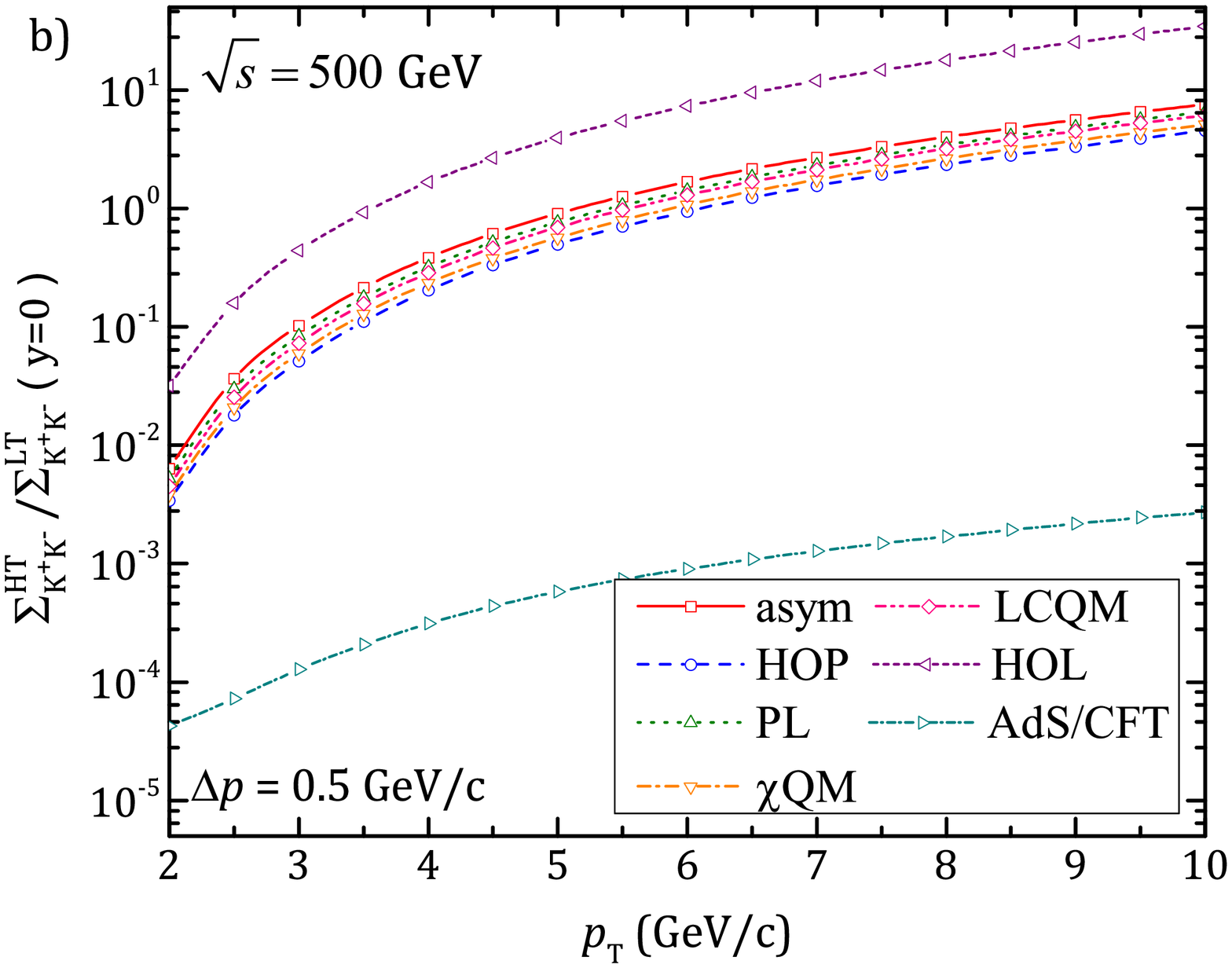}
\includegraphics[scale=0.40]{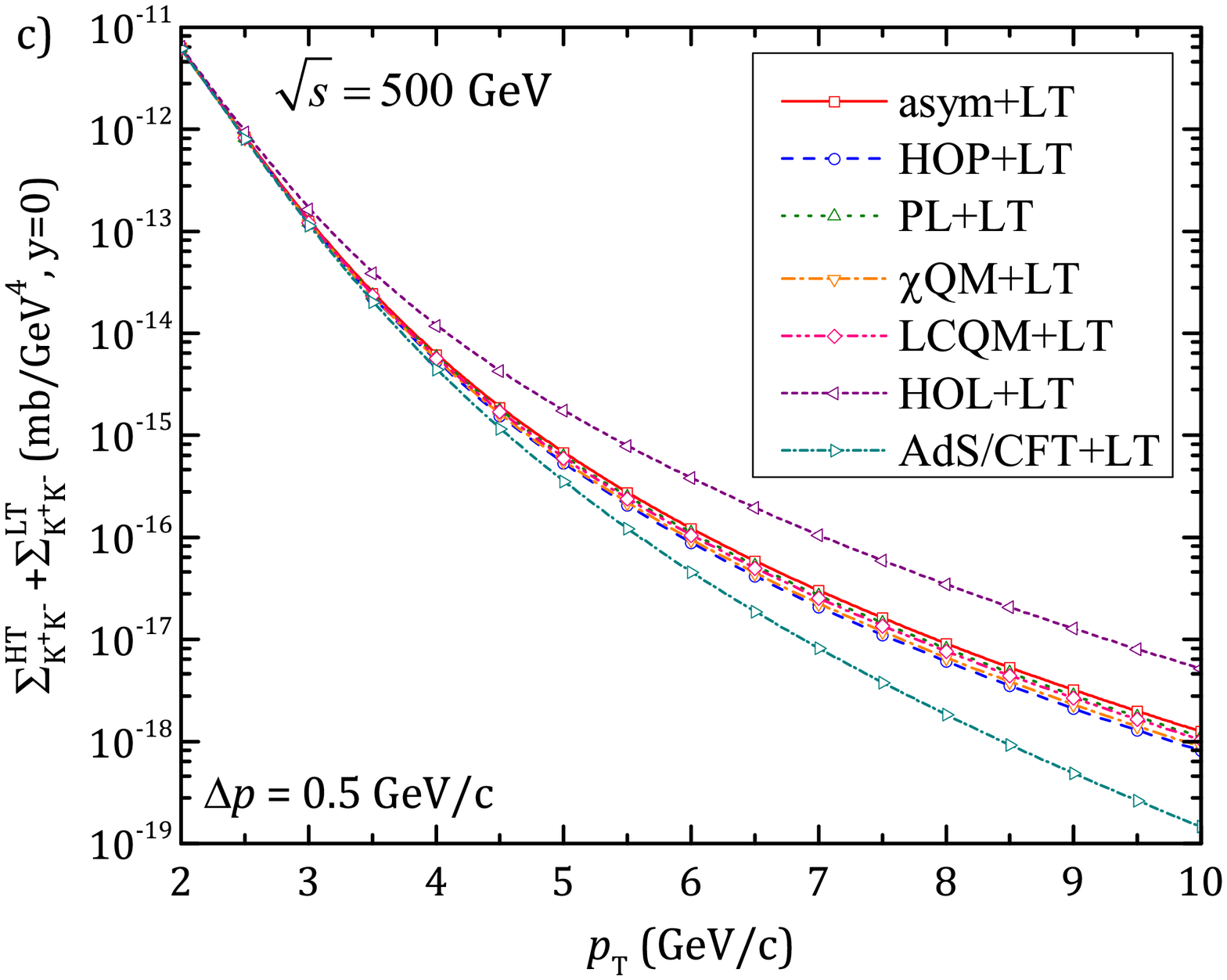}
     \end{center}
\caption{a) LT and HT contributions to charged-kaon pair production $p\bar {p}\to K^{+}K^{-}X$,
b) ratio of HT to LT and c) sum of these contributions as a function of the transverse momentum $p_{T}$
for momentum cut-off parameter $\Delta p=0.5$ GeV/c at $\sqrt s=500$ GeV.
The insert figure in (a) shows HT and LT contributions for a interval of $p_{T}$ from 3.5 to 4.5 GeV/c.}
\label{fig:fig6}
\end{figure*}
To have a quantitative understanding of the effects of HT contributions on the charged kaon pair production,
it is convenient to compute the ratio of HT to LT contributions, namely, $ \Sigma_{K^+K^-}^{HT}/ \Sigma_{K^+K^-}^{LT}$.
We have examined the dependence of HT and LT contributions to charged kaon pair production, their sum and
their ratio on the transverse momentum $p_{T}$, the rapidity $y$ of kaon pairs, and the variable $x_T$
for seven different DAs predicted by light-cone formalism, the light-front quark model, the nonlocal chiral quark model and
the light-front holographic AdS/CFT approach.

\begin{figure*}[!ht]
    \begin{center}
\includegraphics[scale=0.40]{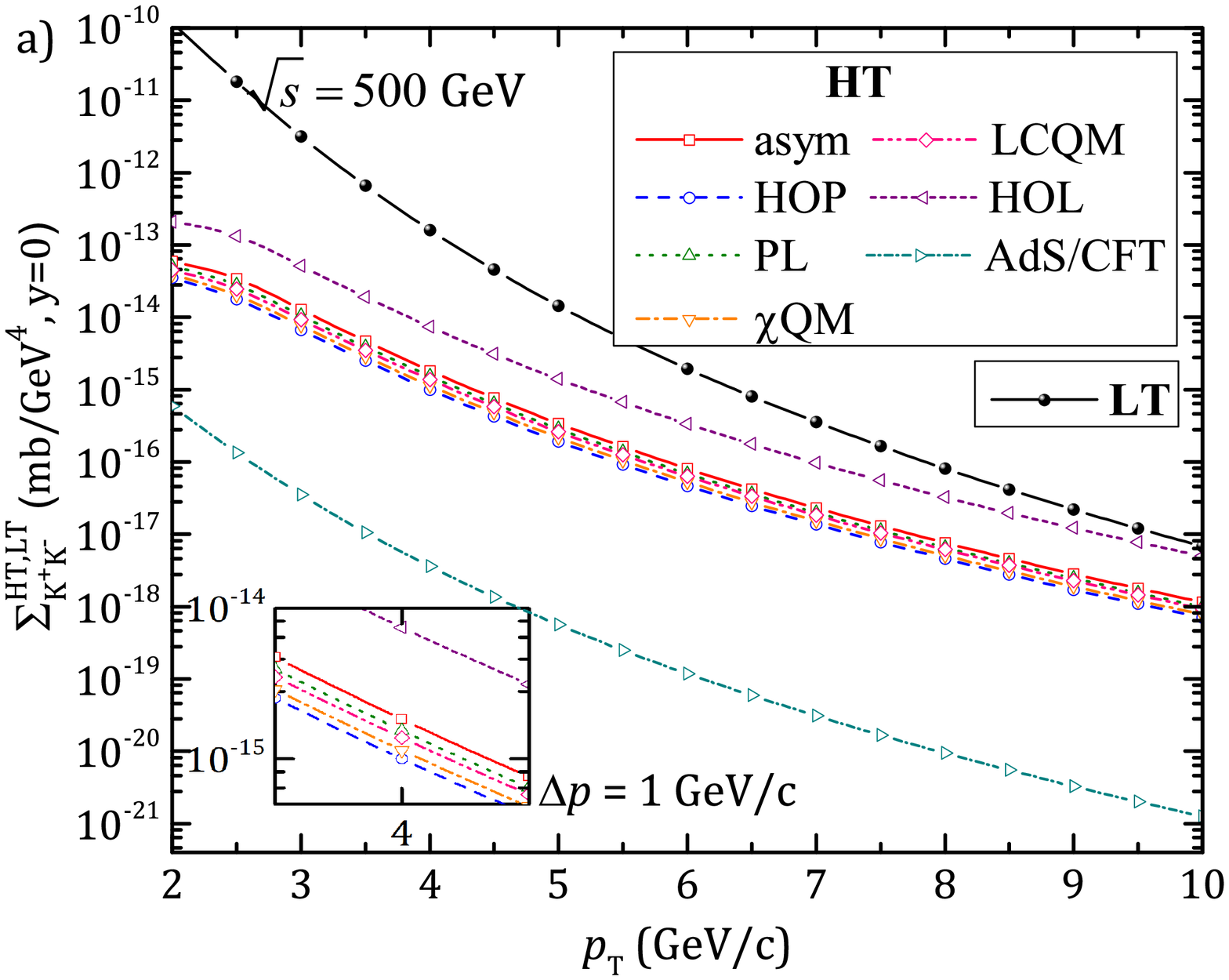}
\includegraphics[scale=0.40]{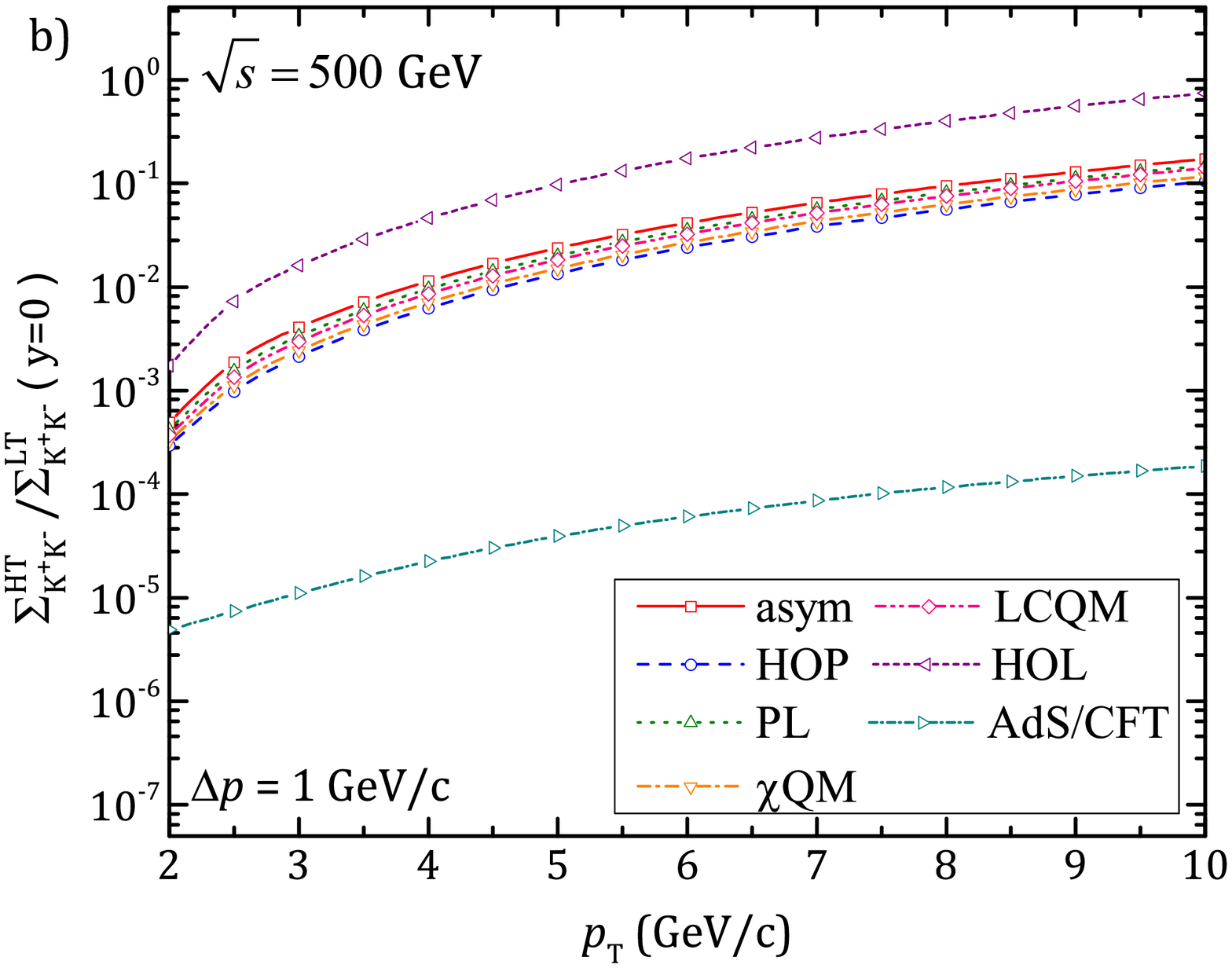}
\includegraphics[scale=0.40]{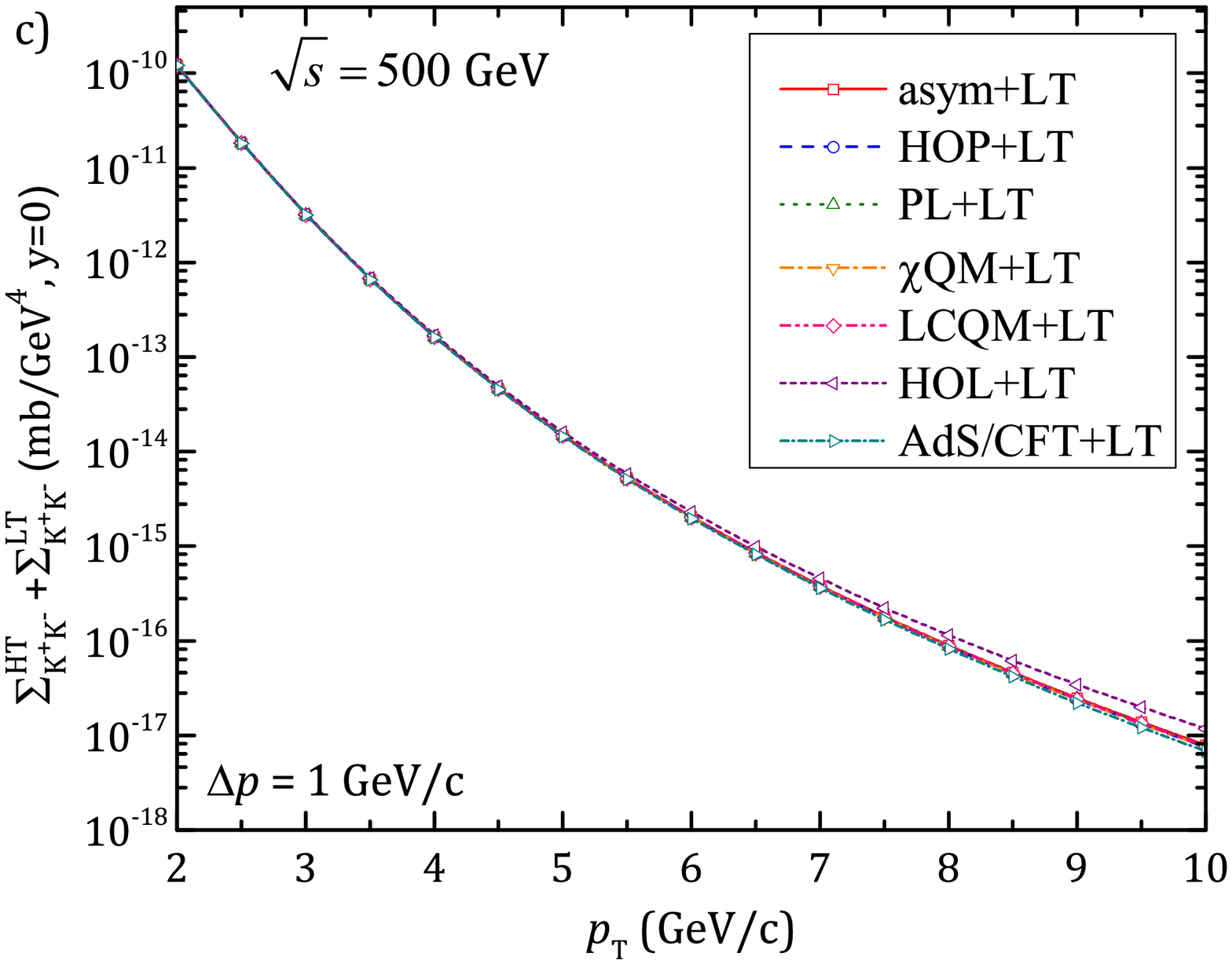}
     \end{center}
\caption{The same as in Fig.~\ref{fig:fig6}, but for $\Delta p=1 \gevc$.}
\label{fig:fig7}
\end{figure*}
We plot the dependence of the HT, LT contributions, ratio of HT to LT and sum of HT and LT on the transverse momentum $p_{T}$ ranging from 2 to 10 GeV/c
at the center-of-mass energy  $\sqrt s=500$ GeV with rapidities of kaons $y=y_1=y_2=0$ for momentum cut-off parameter $\Delta p=0.5$ GeV/c in Fig.~\ref{fig:fig6} and $\Delta p=1$ GeV/c in Fig.~\ref{fig:fig7}.
We do not compute HT and LT contributions for $p_T<2 \gevc$, however, since the theory of perturbation becomes increasingly less reliable in
that region.
Both LT and HT cross sections decrease smoothly with increasing the transverse momentum for each DAs.
It is clear that HT contributions depend on the choice of different kaon DAs.
Note that the DAs of LCQM, HOP, PL, and $\chi$QM give results which are close in shape to those for the asymptotic DA,
but $\Sigma_{K^+K^-}^{HOL}$ is larger and $\Sigma_{K^+K^-}^{AdS/CFT}$ is smaller than them by one and three orders of magnitude, respectively.
The HT contribution calculated for HOL is roughly $\% 77$, $\% 81$, $\% 82$,  $\% 85$, $\% 87$ and four order of magnitude
larger than those for asym, PL, LQCM, $\chi$QM, HOP and AdS/CFT, respectively.
In other words, the HT contributions are sorted in descending order according to
our DAs as $\Sigma_{K^+K^-}^{HOL}> \Sigma_{K^+K^-}^{asym}> \Sigma_{K^+K^-}^{PL}> \Sigma_{K^+K^-}^{LCQM}> \Sigma_{K^+K^-}^{\chi QM}
> \Sigma_{K^+K^-}^{HOP}> \Sigma_{K^+K^-}^{AdS/CFT}$. In particular, the
HT cross section in HOL, $\Sigma_{K^+K^-}^{HOL}$,
reaches a value of $1.4\times 10^{-15}\mb$ for both values of $\Delta p$ at $p_T=5\gevc$,
while value of LT is $3.5\times 10^{-16}\mb$ for $\Delta p=0.5$ GeV/c
and $1.4\times 10^{-14}\mb$ for $\Delta p=1 \gevc$.

The ratio of HT to LT contributions will allow us to determine such regions in the phase space where HT contributions are essentially observable.
The ratio,  $\Sigma_{K^+K^-}^{HT}/\Sigma_{K^+K^-}^{LT}$, increases systemically with increasing
the transverse momentum, because $z_{min}$ comes closer to 1 and thus $\Sigma_{K^+K^-}^{LT}$ decreases.
Figure~\ref{fig:fig6}(b) shows that  $\Sigma_{K^+K^-}^{HT}/\Sigma_{K^+K^-}^{LT}$ at $\Delta p=0.5$ GeV/c increases from 0.006 to 7.57 for asymptotic,
0.003 to 4.55 for HOP, 0.005 to 6.56 for the PL, 0.004 to 5.09 for $\chi$QM,
0.004 to 6.12 for the LCQM, 0.032 to 34.54 for HOL and $4\times 10^{-5}$ to 0.003 for AdS/CFT when $p_T$ runs from 2 to 10 GeV/c.
Particularly, HT contributions become significant at $p_T\geqslant5.5 \gevc$ for asym, HOP, PL, $\chi$QM and LCQM, and $p_T\geqslant3.5$ GeV/c for HOL.
Figure~\ref{fig:fig7}(b) displays that  $\Sigma_{K^+K^-}^{HT}/\Sigma_{K^+K^-}^{LT}$ at $\Delta p=1 \gevc$ increases from $5\times 10^{-4}$ to 0.17 for asymptotic,
$3\times 10^{-4}$ to 0.10 for HOP, $4\times 10^{-4}$ to 0.15 for the PL, $3\times 10^{-4}$ to 0.12 for $\chi$QM,
$4\times 10^{-4}$ to 0.14 for the LCQM, 0.002 to 0.74 for HOL and $5\times 10^{-6}$ to $2\times 10^{-4}$ for AdS/CFT when $p_T$ runs from 2 to 10 GeV/c.

\begin{table*}[ht]
\caption{The individual HT contributions from direct and semi-direct hard-scattering processes at $p_T=5 \gevc$ for each DAs,
where all contributions are given in $\mb$.}
\label{tab:table3}
\begin{ruledtabular}
\begin{tabular}{lccccR[.][.]{3}{3}}
&\multicolumn{2}{c}{\textbf{Direct production}}&\multicolumn{2}{c}{\textbf{Semi-direct production}}&\multicolumn{1}{c}{\textbf{Ratio of HT to LT}}\\ \cline{2-6}
\textbf{DAs} &\multicolumn{1}{c}{$\text{g}\text{g}\to K^+K^-$}&\multicolumn{1}{c}{$q\bar{q}\to K^+K^-$}
&\multicolumn{1}{c}{$q\bar{q}' \to K^\pm \text{g}$}&\multicolumn{1}{c}{ $q\text{g}\to K^{\pm}q' + \bar{q} \text{g}\to K^{\pm}\bar{q}'$ }&\multicolumn{1}{c}{$\Sigma_{K^+K^-}^{HT}/\Sigma_{K^+K^-}^{LT}$}\\
\hline
\textbf{asym}        &3.06$\times10^{-15}$  &7.64$\times10^{-19}$  &3.46$\times10^{-19}$  &1.20$\times10^{-17}$ &0.9065\\
\textbf{HOP}         &1.65$\times10^{-16}$  &3.44$\times10^{-19}$  &2.70$\times10^{-19}$  &9.36$\times10^{-18}$ &0.4976\\
\textbf{PL}          &2.57$\times10^{-16}$  &5.91$\times10^{-19}$  &3.22$\times10^{-19}$  &1.12$\times10^{-17}$ &0.7644\\
\textbf{$\chi$QM}    &1.89$\times10^{-16}$  &3.45$\times10^{-19}$  &2.83$\times10^{-19}$  &9.83$\times10^{-18}$  &0.56729\\
\textbf{LCQM }       &2.32$\times10^{-16}$  &3.86$\times10^{-19}$  &3.10$\times10^{-19}$  &1.07$\times10^{-17}$ &0.69093\\
\textbf{HOL }        &1.29$\times10^{-15}$  &7.02$\times10^{-17}$  &6.16$\times10^{-19}$  &2.14$\times10^{-17}$ &4.09628\\
\textbf{AdS/CFT }    &1.74$\times10^{-21}$  &2.25$\times10^{-24}$  &5.65$\times10^{-21}$  &1.96$\times10^{-19}$ &0.0005779\\
\end{tabular}
\end{ruledtabular}
\end{table*}

\begin{figure*}[!ht]
    \begin{center}
\includegraphics[scale=0.40]{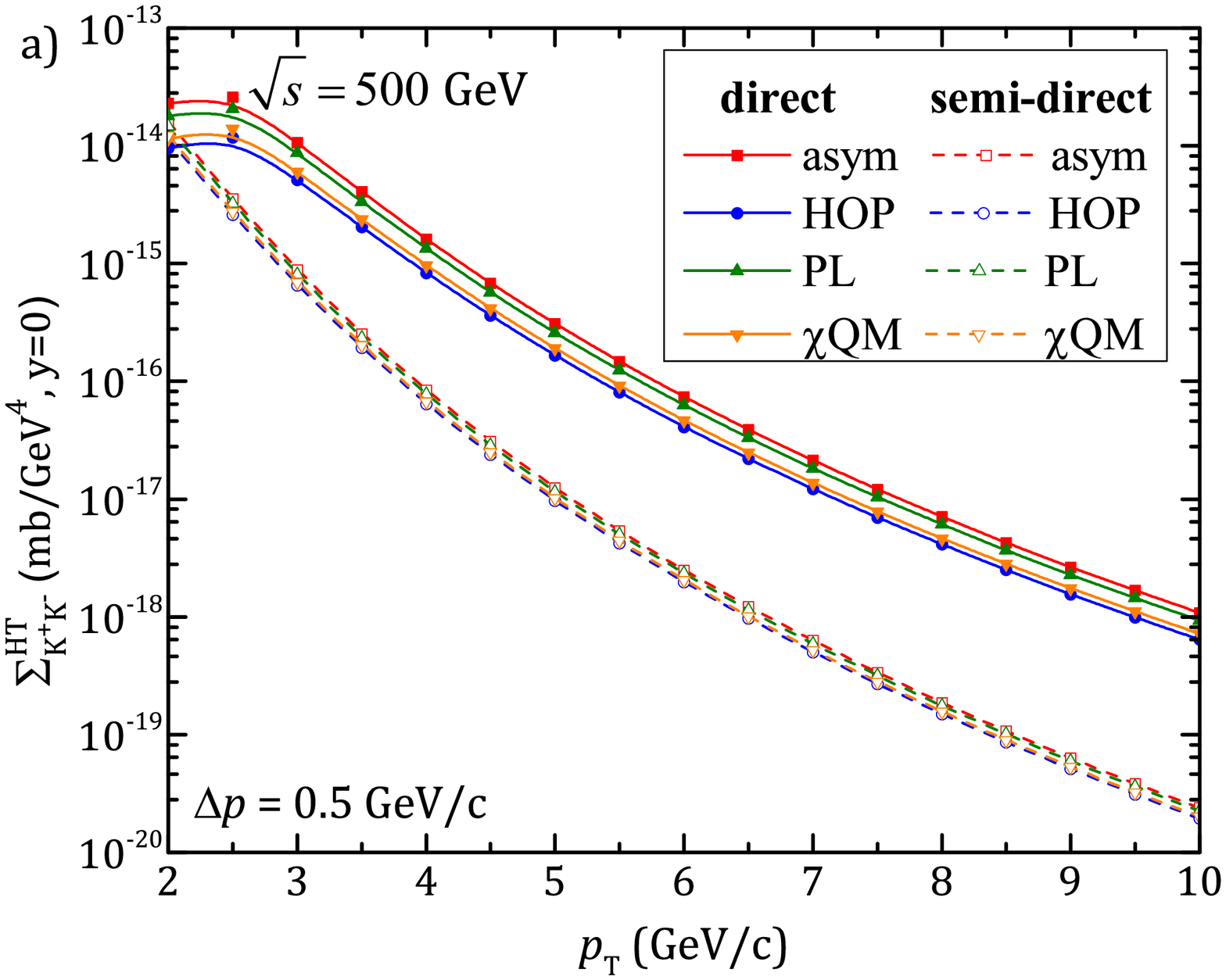}
\includegraphics[scale=0.40]{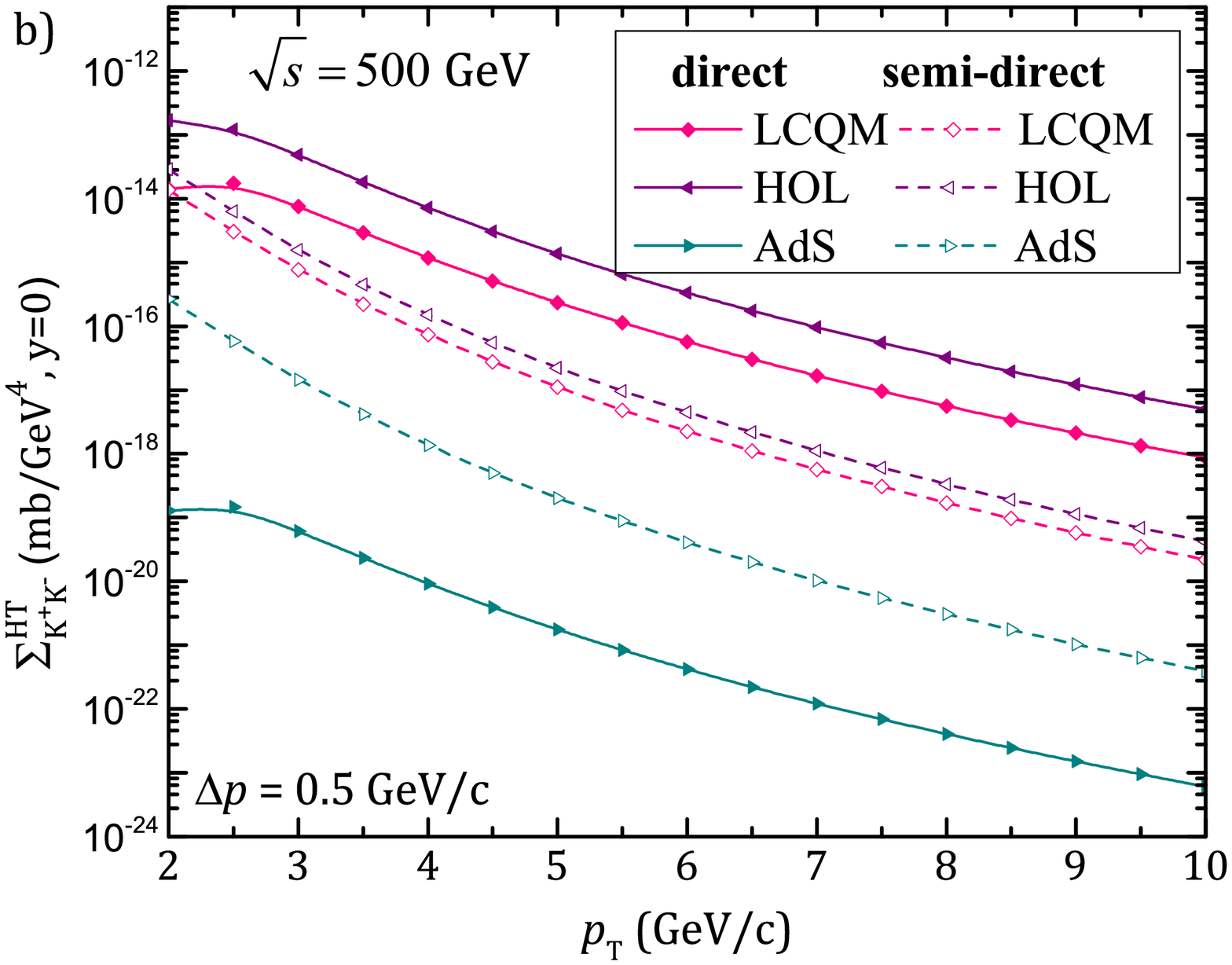}
     \end{center}
\caption{a) HT contributions from direct and semi-direct production for a) asym, HOP, PL, and $\chi$QM,
b) LCQM, HOL and AdS/CFT as a function of the transverse momentum $p_{T}$ for momentum cut-off parameter $\Delta p=0.5$ GeV/c at $\sqrt s=500$  GeV.
The solid lines and dashed-lines indicate to HT contributions from direct-production and semi-direct production, respectively.}
\label{fig:fig8}
\end{figure*}
These results show that the ratio $\Sigma_{K^+K^-}^{HT}/\Sigma_{K^+K^-}^{LT}$ is mostly sensitive
according to variation of the $\Delta p$ and $p_T$.
For small value of $\Delta p$, it reaches considerably larger values.
The corresponding ratio, on the other hand, increases by about two orders of magnitude for $\Delta p=1 \gevc$
and three orders of magnitude for $\Delta p=0.5 \gevc$ when the transverse momentum $p_T$ varies from 2 to 10 GeV/c for each DA.
The ratio of HT contributions calculated with $\Delta p=0.5 \gevc$ and $\Delta p=1 \gevc$,
are constant within around $\% 93$ for asym, HOP, $\chi$QM, LQCM, $\% 99$ for HOL and $\% 35$ for AdS/CFT between $p_T=3$ and 10 GeV/c.
The LT cross section calculated with $\Delta p=1 \gevc$ are, however,
around $\% 97$ larger than one calculated with $\Delta p=0.5 \gevc$.

It is interesting to see what the relative contributions are of the different internal mechanisms.
In Fig.~\ref{fig:fig8}, we present the dependence of the HT contributions from direct and semi-direct production, separately,
on the transverse momentum $p_{T}$ at $\sqrt s=500$ GeV for each DA.
For direct kaon pair production, hard-scattering subprocesses are $\text{g}\text{g}\to K^+K^-$ and
$q\bar{q}\to K^+K^-$ for $q=u$ and $s$. For semi-direct kaon pair production, hard-scattering subprocesses are $q\bar{q}' \to K^\pm \text{g}$ and
$q\text{g}\to K^{\pm}q'$ (and $\bar{q} \text{g}\to K^{\pm}\bar{q}'$).
Direct production processes make dominant contributions for asym, HOP, PL, $\chi$QM, LCQM and HOL,
while for AdS/CFT, semi-direct production processes have dominant contribution.
When $p_T$ runs from 2 to 10 GeV/c, HT contributions from direct production decrease
by around four orders of magnitude for asym, HOP, PL, $\chi$QM and LCQM, and five orders of magnitude for HOL and AdS/CFT.
Also, the HT contributions from semi-direct production decrease by about six orders of magnitude for all DAs.
For example, for HOL, HT contributions decreases from $1.67\times 10^{-13}$ to $4.98\times 10^{-18}\mb$ in direct production
and $2.88\times 10^{-14}$ to $4.23\times 10^{-20}\mb$ in semi-direct production.
However, it should be emphasized that the difference between HT contributions from direct and semi-direct productions increase
with increments of the transverse momentum $p_{T}$.

Moreover, with a view to make easy precise comparisons with the
experimental results, we list individual HT contributions from direct and semi-direct hard-scattering processes at $p_T=5 \gevc$
for each DAs in Table~\ref{tab:table3}. It is seen from this table that direct production HT contributions are dominated by the process $\text{g}\text{g}\to K^+K^-$ as expected.
Semi-direct production HT contributions are dominated by the process $^(\bar{q}^)\text{g}\to K^{\pm}$$^(\bar{q}^)$$'$.

\begin{figure*}[!ht]
    \begin{center}
\includegraphics[scale=0.40]{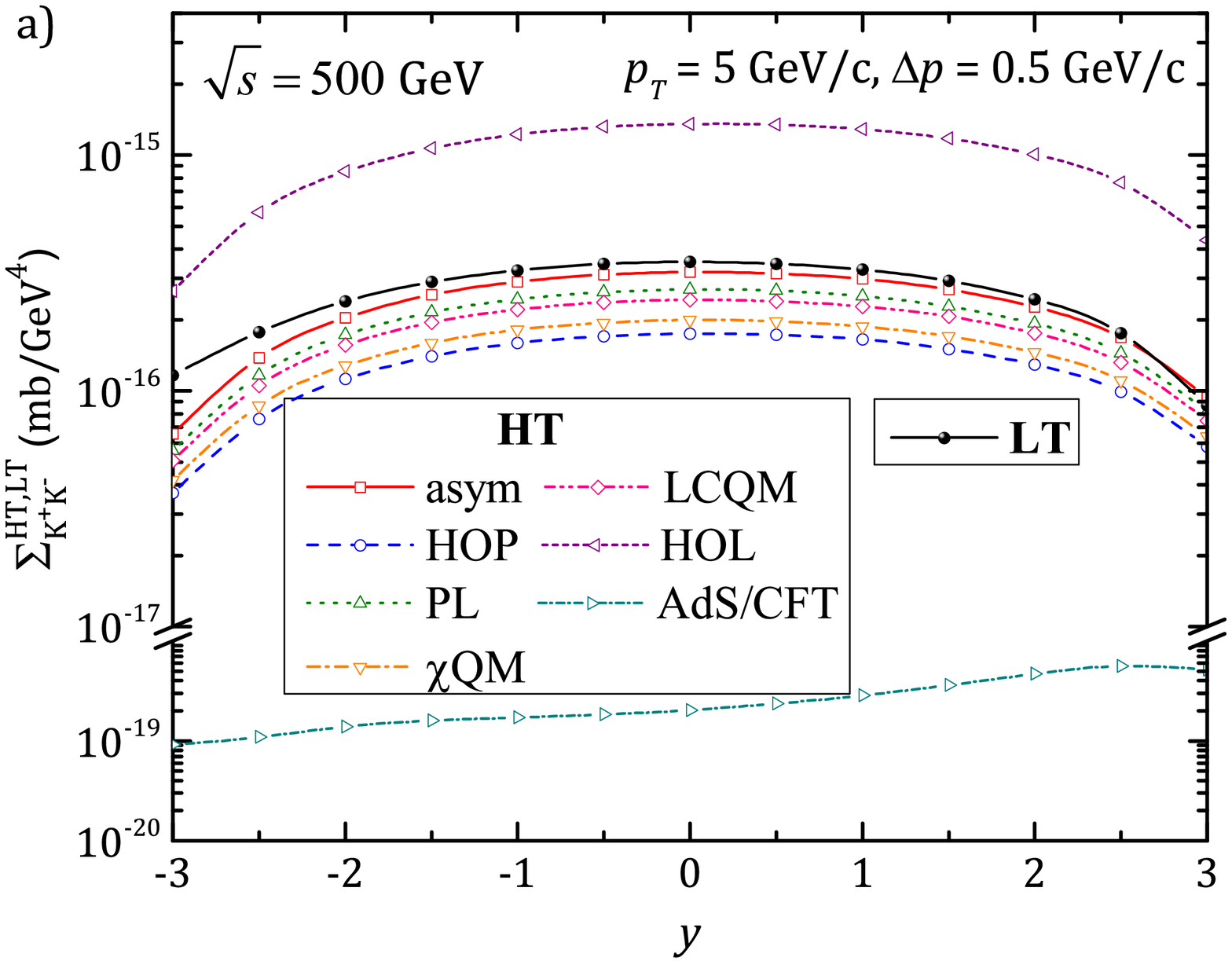}
\includegraphics[scale=0.40]{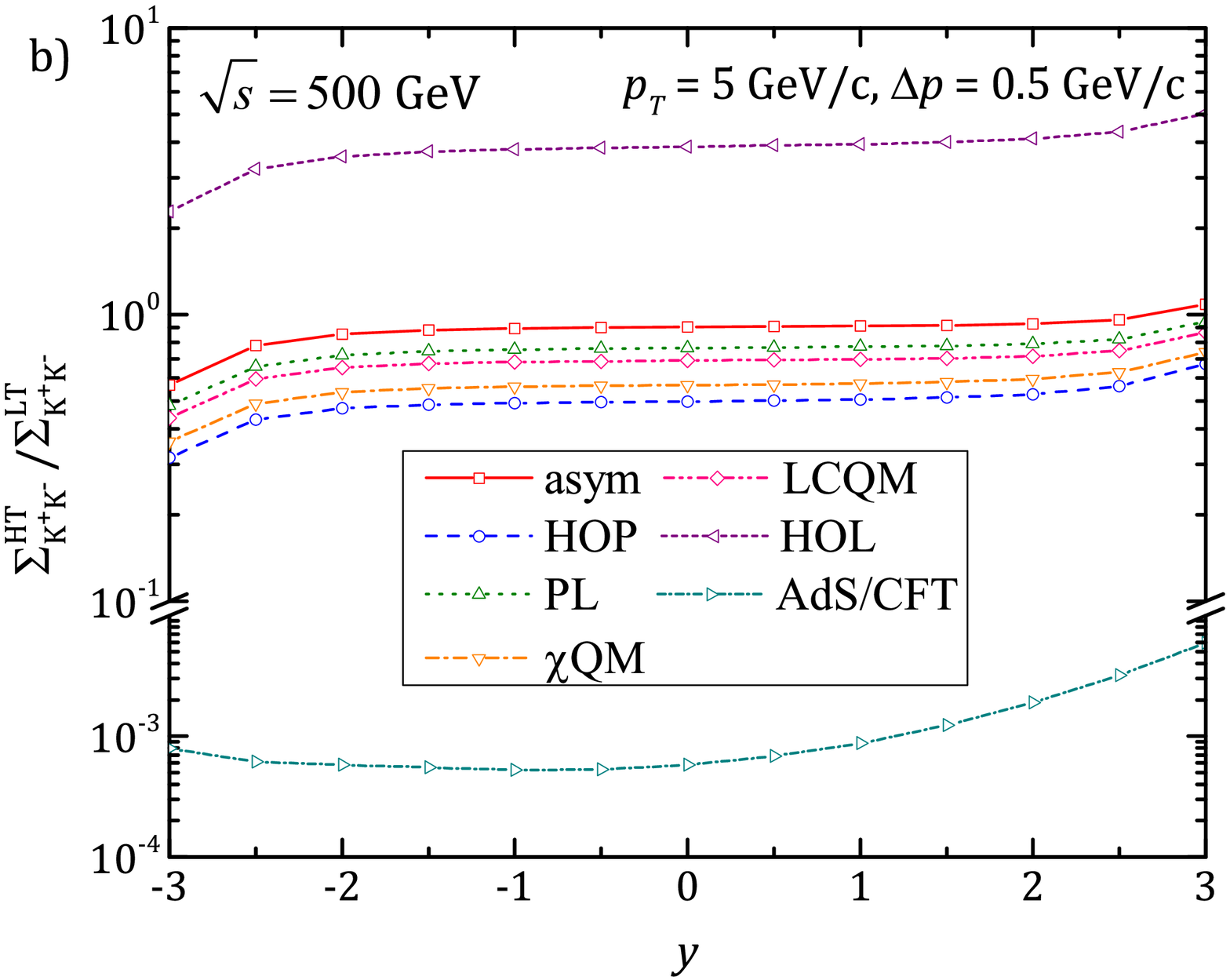}
\includegraphics[scale=0.40]{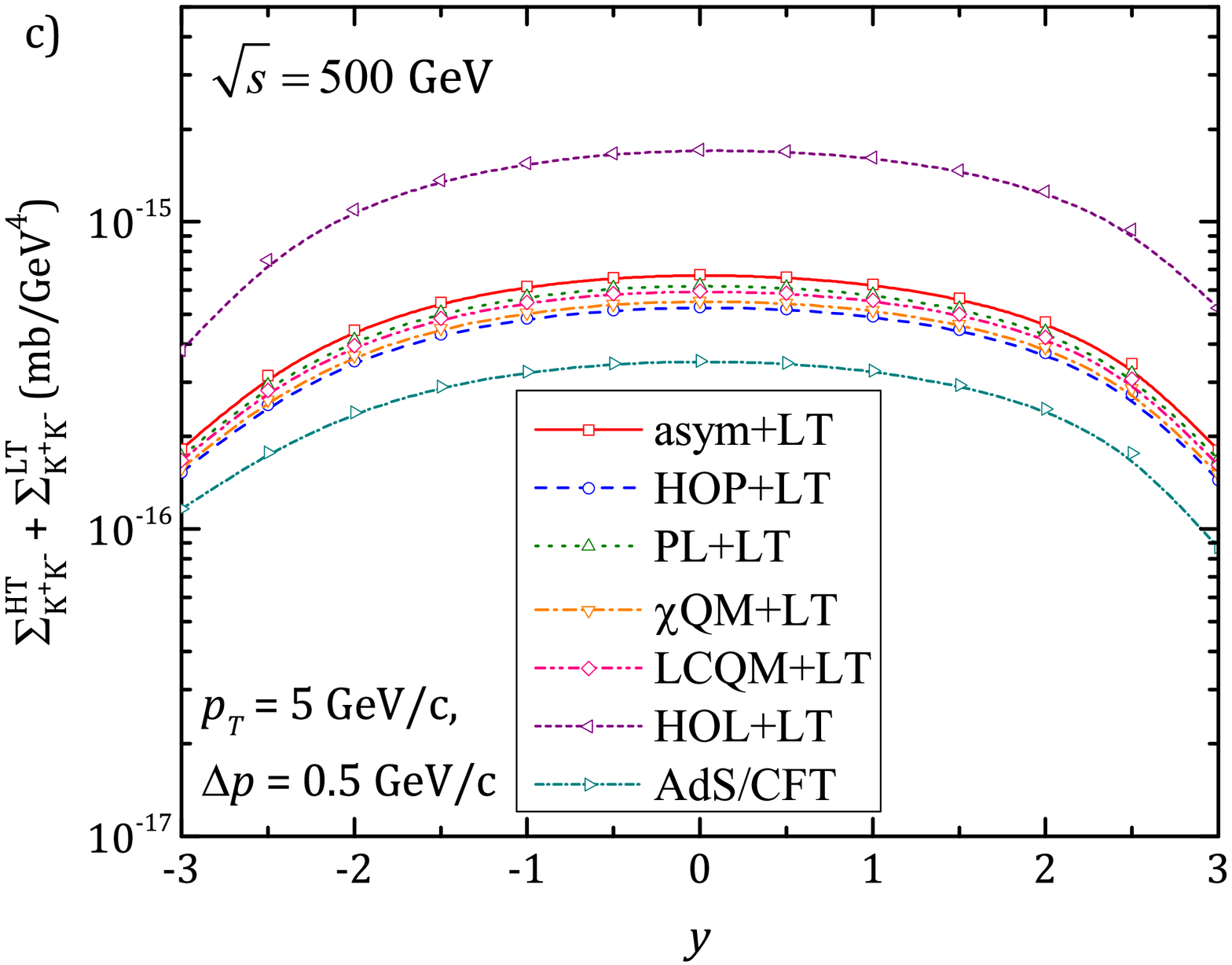}
     \end{center}
\caption{a) LT and HT contributions to charged-kaon pair production $p\bar {p}\to K^{+}K^{-}X$,
b) ratio of HT to LT and c) sum of these contributions as a function of the rapidity $y=y_1=y_2$ of kaon pairs
for momentum cut-off parameter $\Delta p=0.5$ GeV/c and $p_T=5 \gevc$ at $\sqrt s=500$ GeV.}
\label{fig:fig9}
\end{figure*}
We exhibit the dependence of the HT, LT contributions, ratio of HT to LT and sum contribution on the rapidity $y=y_1=y_2$ of kaon pairs
varied in the range from -3 to 3 at $\sqrt s=500$ GeV for each DA in Fig.~\ref{fig:fig9}. The rapidity distribution demonstrates
the same dominant contributions in view of DAs as the ones in the transverse momentum dependence of the cross section.
It is seen that the HT and LT cross sections for all DAs of kaons except AdS/CFT, have a maximum at the point y=0.
Additionally, they are almost symmetric according to y=0.
However, the HT contributions in the region of positif rapidity are always somewhat larger than those in region of negative rapidity.
When the rapidity goes up from 0 to 3, the HT cross sections decrease by around
$\% 70$ for asym, HOP, PL, $\chi$QM, LCQM and HOL,
while grows by a factor of 1.5 for AdS/CFT. The LT contribution also decreases by $\% 75$.
The ratio of HT to LT contributions is getting bigger slowly when $y$ goes up from -3 to -2
and remains almost stable in a interval of rapidity from -2 to 2 and
then continues to increase slowly with increments of rapidity from 2 to 3 for all DAs of kaons except AdS/CFT.

\begin{figure*}[!ht]
    \begin{center}
\includegraphics[scale=0.40]{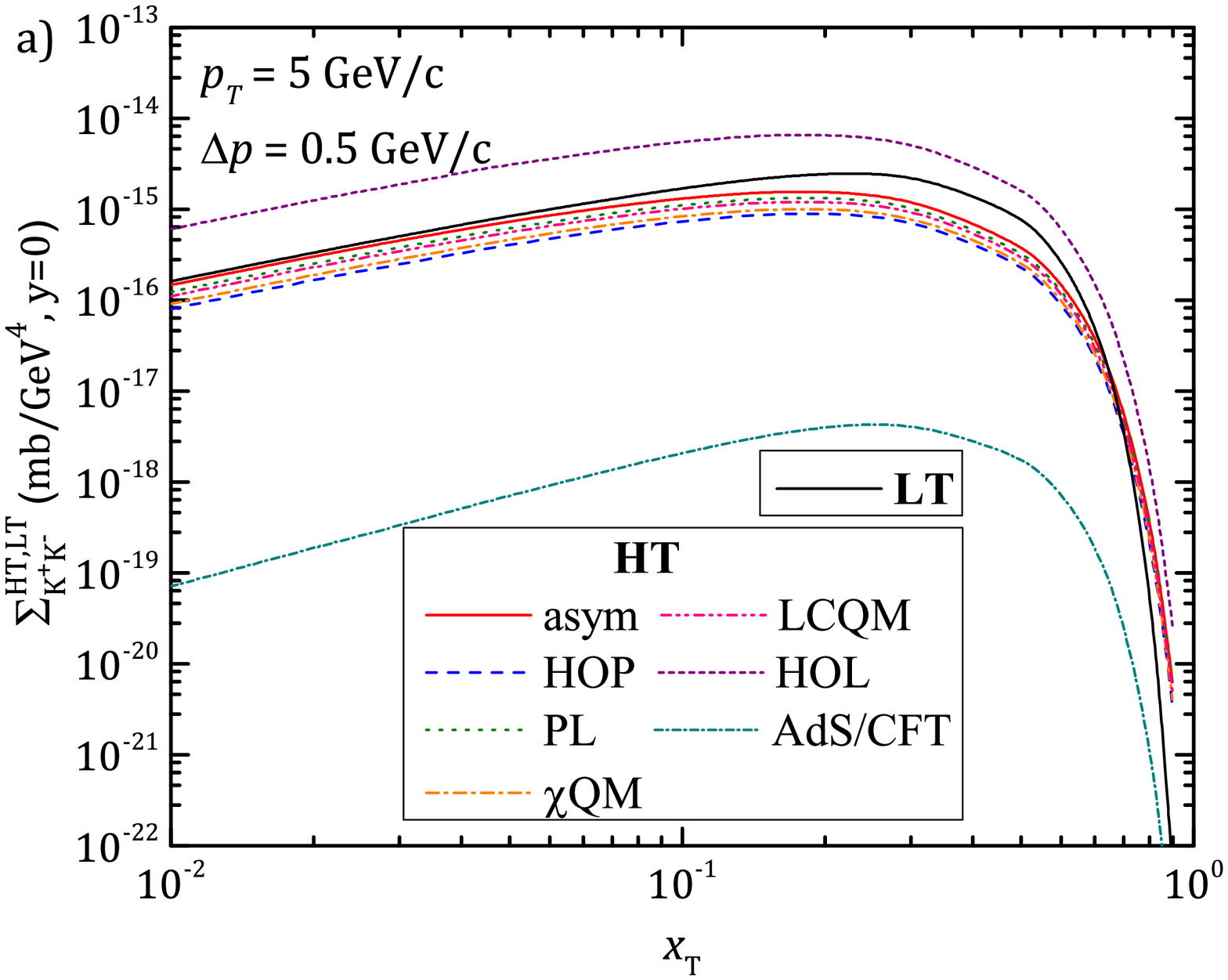}
\includegraphics[scale=0.40]{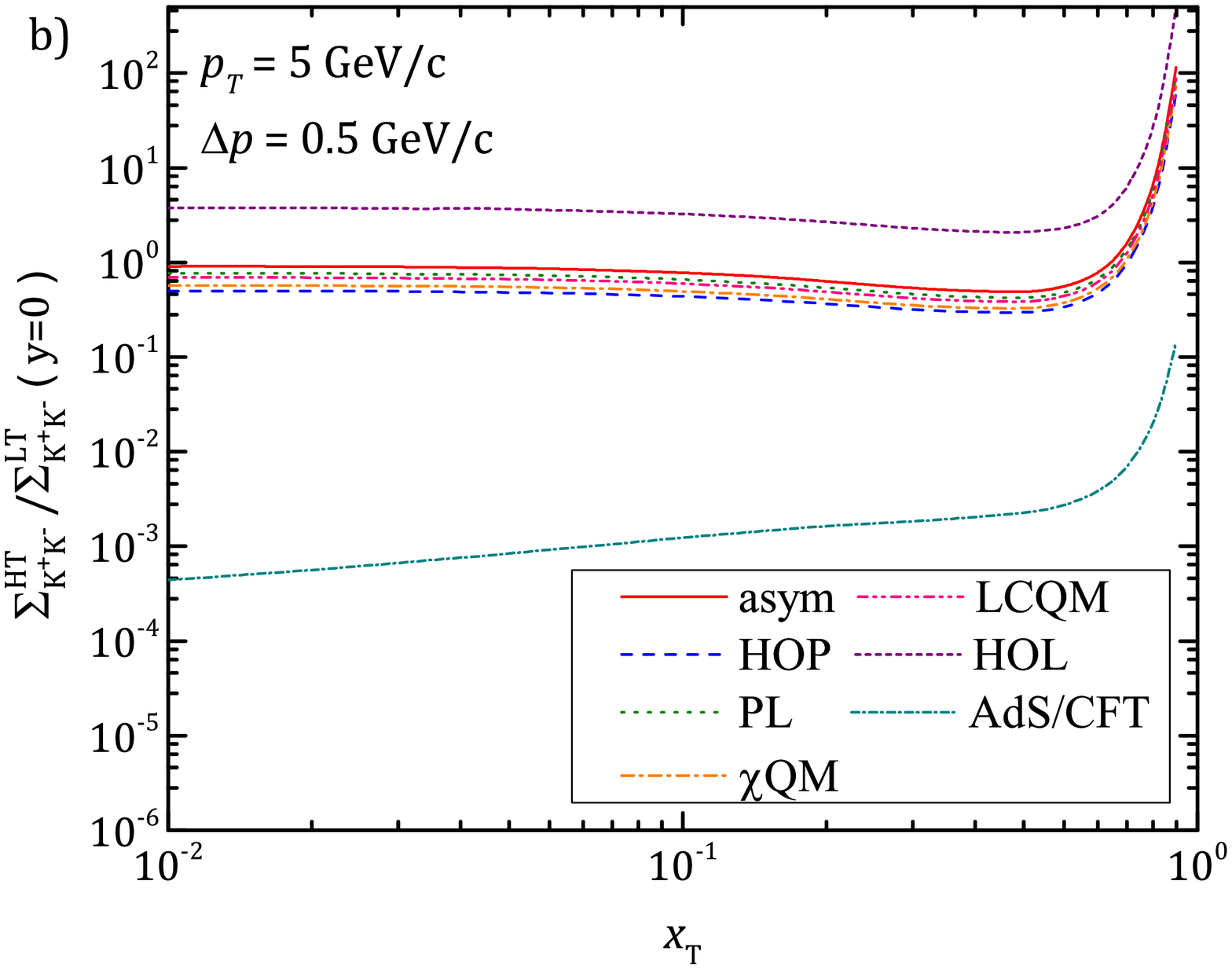}
     \end{center}
\caption{a) LT and HT contributions to charged-kaon pair production $p\bar {p}\to K^{+}K^{-}X$ and
b) ratio of HT to LT contributions as a function of the variable $x_T$ for momentum cut-off parameter $\Delta p=0.5 \gevc$ at $p_T=5\gevc$.}
\label{fig:fig10}
\end{figure*}
For different center of mass energies, the HT and LT differential cross sections are constructed and
compared to a scaling with the variable $x_T=2 p_T/\sqrt{s}$.
We have performed the above numerical results at $\sqrt{s}=500$ GeV.
To compared with other energies we also show the dependence of HT, LT contributions
and ratio of HT to LT on the variable $x_T$ ranging from $10^{-2}$ to $10^{0}$ at the $p_{T}=5 \gevc$
with rapidities of kaons $y=y_1=y_2=0$ for momentum cut-off parameter $\Delta p=0.5$ GeV/c in Figs.~\ref{fig:fig10}(a)-(b).
These plots reveal that the distribution of variable $x_T$ also demonstrates the same dominant contributions in view of DAs
as the ones in the transverse momentum dependence of the cross section.
Both HT and LT contributions increase slowly when $x_T$ goes up from $0.01$  to 0.2 and then decrease rapidly
with increments of $x_T$ from $0.2$  to $1$ for all DAs of kaons.
Note that the decrease in the contributions is fast since the $x_T$ is in the vicinity of 1, namely, $\sqrt{s} \sim 2 p_T$.
The ratio of HT to LT contributions remain almost stable in a large interval of $x_T$.
This means that the ratio is less sensitive according to varying the center-of-mass energy.

\section{Summary and Prospects}\label{sec:conc}
In this work, the HT contributions, which are included the direct and semi-direct productions of the hard scattering process,
to large-$p_T$ kaon pair production in $p\bar{p}$ collisions have been discussed and
the dependence of HT contributions on kaon-DAs predicted by light-cone formalism, the light-front quark model,
the nonlocal chiral quark model and the light-front holographic AdS/CFT approach have been addressed.

It is observed that the results are significantly depend on the DAs of kaon, and can, hence,
be used for their research.
The basic size of the HT cross sections is seen to differ by several orders of magnitude depending
on the choice of DAs of the produced kaons. The DAs of LCQM, HOP, PL, and $\chi$QM give results
which are close in shape to those for the asymptotic DA,
whereas HT contributions for HOL are larger than them by one order of magnitude
and for AdS/CFT are smaller by three orders of magnitude.

The ratio of HT to LT contributions allows us to determine such regions in the phase space
where HT contributions are essentially observable.
This ratio is sensitive to the transverse momentum $p_T$ and the momentum cut-off parameter $\Delta p$ which is the detection limit for accompanying particles.
For a small value of $\Delta p$ and a large value of $p_T$, HT contributions yield considerably larger values. While the HT effect
on cross section is small at the low $p_T$ region, its effect becomes significant at the large $p_T$ region compared to the LT contribution.

It is obvious that total contribution of direct production hard-scattering processes is larger than ones of semi-direct production processes in most cases.
The HT process $\text{g}\text{g}\to K^+K^-$ gives the largest contribution to the inclusive cross section at large $p_T$ for all DAs.
However, among semi-direct production processes, the process  $q\text{g}\to K^{\pm}q'$ or $\bar{q} \text{g}\to K^{\pm}\bar{q}'$ has dominant contributions for all DAs.

The rapidity distribution exhibits the same dominant contributions in view of DAs as the ones in
the transverse momentum dependence of the cross section. The HT contributions are enhanced in the region of positive rapidity.

Consequently, we can point out that the HT processes for large-$p_T$ kaon pair production have a non-negligible contribution, where the
kaons are produced directly in the hard-scattering subprocess, rather than by gluon and quark fragmentation.
Inclusive kaon pair production provides a significant test case in which HT contributions dominate those of
LT in certain kinematic regions.
The HT contributions can be used to theoretical interpretation of the future experimental data
for the charged kaon pair production in $p\bar{p}$ collisions. The results of this work will be helpful
to providing a basic test of the short distance structure of QCD as well as to determine more precise DAs of kaon.

\begin{acknowledgments}
M. Demirci is grateful to T. M. Aliev for useful discussions. A. I. Ahmadov is grateful for the financial support by the Science Development Foundation under the President of the Republic of Azerbaijan-Grant no: EIF/MQM/Elm-Tehsil-1-2016-1(26)-71/11/1. We have drawn the corresponding Feynman diagrams with the help of the program~\texttt{JaxoDraw}~\cite{JaxoDraw}.
\end{acknowledgments}

\end{document}